\documentclass[%
 reprint,
superscriptaddress,
 amsmath,amssymb,
aps,
pra,
prstper,
floatfix,
]{revtex4-2}
\usepackage{setspace} 
\usepackage{graphicx}% Include figure files
\usepackage{dcolumn}% Align table columns on decimal point
\usepackage{xcolor}
\usepackage{bm}% bold math
\usepackage{float}
\usepackage[%  
    colorlinks=true,
    pdfborder={0 0 0},
    linkcolor=blue
]{hyperref}
\usepackage{comment}
\usepackage{amsmath}
\usepackage{ulem} % pour barrer des trucs (YC)  commande \sout

\usepackage{siunitx}
\begin{document}

\preprint{APS/123-QED}

\title{Cavity quantum electrodynamics with single perovskite quantum dots}% Force line breaks with \\
\author{Zakaria Said}
\affiliation{%
 Laboratoire de Physique de l’Ecole normale supérieure, ENS, Université PSL, CNRS, Sorbonne Université, Université Paris Cité, F-75005 Paris, France\\
}%
 %\altaffiliation[Also at ]{Laboratoire de Physique de l’ENS, Université PSL, CNRS, Sorbonne Université, Université Paris Cité, 75005 Paris, France}
 \author{Marina Cagnon Trouche}
 \affiliation{%
 Laboratoire de Physique de l’Ecole normale supérieure, ENS, Université PSL, CNRS, Sorbonne Université, Université Paris Cité, F-75005 Paris, France\\
}%

 \author{Antoine Borel}
 \affiliation{%
 Laboratoire de Physique de l’Ecole normale supérieure, ENS, Université PSL, CNRS, Sorbonne Université, Université Paris Cité, F-75005 Paris, France\\
}%

 %\altaffiliation[]{}
 \author{Mohamed-Raouf Amara}
 \affiliation{%
 Laboratoire de Physique de l’Ecole normale supérieure, ENS, Université PSL, CNRS, Sorbonne Université, Université Paris Cité, F-75005 Paris, France\\
}%
 %\altaffiliation[Also at ]{Physics Department, XYZ University.}

 %\altaffiliation[Also at ]{Physics Department, XYZ University.}
  \author{Jakob Reichel}
  \affiliation{%
Laboratoire Kastler Brossel, Sorbonne Université, CNRS, ENS - Université PSL, Collège de France, Paris F-75252, France\\
}

\author{Christophe Voisin}
\affiliation{%
 Laboratoire de Physique de l’Ecole normale supérieure, ENS, Université PSL, CNRS, Sorbonne Université, Université Paris Cité, F-75005 Paris, France\\
}%

\author{Carole Diederichs}%
 \email{carole.diederichs@phys.ens.fr}
 
\affiliation{%
 Laboratoire de Physique de l’Ecole normale supérieure, ENS, Université PSL, CNRS, Sorbonne Université, Université Paris Cité, F-75005 Paris, France\\
}%
\affiliation{Institut Universitaire de France (IUF), 75231 Paris, France}
 %\altaffiliation[Also at ]{Physics Department, XYZ University.}
 \author{Yannick Chassagneux}%
 \email{yannick.chassagneux@phys.ens.fr}
 \affiliation{%
 Laboratoire de Physique de l’Ecole normale supérieure, ENS, Université PSL, CNRS, Sorbonne Université, Université Paris Cité, F-75005 Paris, France\\
}%

\date{\today}


\begin{comment}
 \begin{itemize}
    \item Abstract: up to 200 words, unreferenced.
    \item Main text: up to 3000 words, excluding abstract, Methods, references and figure legends.
    \item Article should be divided as follows: Introduction (without heading), Results, Discussion, Online Methods.
    \item Results and Methods should be divided by topical subheadings; the Discussion does not contain subheadings.
\end{itemize}
\end{comment}

%TC:ignore
\begin{abstract}
Quantum emitters of single indistinguishable photons play a key role in quantum technologies. Among condensed matter systems, colloidal perovskite quantum dots have emerged as promising candidates, exhibiting high-purity single photon emission at room temperature and two-photon interference visibilities up to 0.5 at cryogenic temperatures. Achieving deterministic coupling of individual perovskite quantum dots to photonic structures is now a critical step towards harnessing cavity quantum electrodynamics (cQED) effects, such as the Purcell effect, to enhance single photon emission rate and indistinguishability. Here, we demonstrate the deterministic and reversible coupling of individual CsPbBr$_{3}$ perovskite quantum dots to a tunable, high-quality factor, low mode volume fiber-based Fabry-Pérot microcavity at 10~\si{\kelvin}. By spatially and spectrally tuning the cavity mode in resonance with the quantum dot emission, we observe up to a twofold increase in single photon emission rates. We build on the original multiplet excitonic fine structure to assess the vacuum Rabi coupling strength ($g \simeq$ 40~\si{\micro eV}) from the shaping of the spectral profile of the emission upon increasing the electromagnetic confinement. This approach also made it possible to delineate the contributions of spectral diffusion and pure dephasing to the total linewidth of emission, paving the way to a fully optimized control of the emission properties of cavity coupled perovskite quantum dots.
\end{abstract}
%TC:endignore

\maketitle

% INTRODUCTION
Since their first synthesis in 2015 \cite{Protesescu}, colloidal lead halide perovskite quantum dots (pQDs) have emerged as promising nano-emitters for scalable, solution-processed quantum dot-based optoelectronic devices such as photonic sources \cite{sutherland_perovskite_2016}. This is due to their excellent optical properties, including high brightness, tunable optical bandgap, and reduced blinking, achieved through accessible and cost-effective synthesis methods. At the single dot level, pQDs are also among the most promising newcomers in the field of solid-state single-photon emitters \cite{aharonovich_solid-state_2016, hu_superior_2015} showing stable single photon emission with high purity at room temperature \cite{park_room_2015,zhu_room-temperature_2022}. In addition, at cryogenic temperatures, pQDs exhibit long coherence time ($T_2$), comparable to their lifetime ($T_1$) \cite{utzat_coherent_2019, lv_quantum_2019, lv_exciton-acoustic_2021}. This remarkable value has allowed the first demonstration by Kaplan et al. in 2023 of 50\si{\percent} photon indistinguishability in colloidal quantum dots \cite{kaplan_hongoumandel_2023}, whereas perfect photon indistinguishability is achieved in the limit $T_2=2T_1$. Furthermore, phenomena such as radiative biexciton-exciton cascade emission \cite{tamarat_dark_2020, WangOptically}, capable of generating correlated polarized photon pairs, and collective superfluorescence emission in self-assembled pQDs superstructures \cite{raino_superfluorescence_2018}, highlight the assets of perovskite nanostructures for quantum technologies.

To fully exploit these assets, a critical step is to enhance light-matter interaction at the single pQD scale by coupling to photonic structures, such as optical microcavities, to harness cavity quantum electrodynamics (cQED) effects. These effects, whether in the weak or strong coupling regime, depend critically on the spectral and spatial matching between the emitter and the cavity, as well as on the quality factor ($Q$) and mode volume ($V$) of the cavity. In the weak coupling regime, where the vacuum Rabi coupling strength $g$ is smaller than the system losses, the acceleration of the radiative spontaneous emission together with the redirection of the emission in the cavity mode are key features of the Purcell effect \cite{Purcell1946} for the development of efficient single photon sources. In the strong coupling regime, where $g$ exceeds the losses, cQED effects yield few-photon nonlinearities enabling advanced features for quantum photonic logic gates \cite{birnbaum_photon_2005}. Knowing the value of $g$ is therefore crucial to assess the potential of individual quantum emitters for cQED applications. While $g$ is directly imprinted in the spectrum via the Rabi splitting in the strong coupling regime, its evaluation in the weak coupling regime is more challenging as it relies mainly on indirect analysis. Up to now, pQDs have seen limited experimental demonstrations of cavity coupling and the evaluation of $g$ remains unexplored for individual pQDs. Early work demonstrated the coupling of ensembles of pQDs to photonic crystal nanobeams, achieving emission acceleration \cite{yang_spontaneous_2017}. Recently, the coupling of single pQDs to circular Bragg gratings has also been demonstrated with an emission enhancement in the Purcell regime \cite{jun_ultrafast_2023,purkayastha_purcell_2024}. Nevertheless, significant challenges hinder the precise assessment of cQED effects. These include the statistical biases arising from the variations of their local environment when comparing the optical properties of coupled pQDs with those of uncoupled ones. To overcome these limitations, open microcavities offer a compelling solution, especially for randomly positioned quantum emitters, since their flexibility enables deterministic emitter-cavity coupling, as demonstrated for various individual emitters such as carbon nanotubes \cite{jeantetWidelyTunableSinglePhoton2016b,borel_telecom_2023}, NV centers in diamond \cite{riedel_deterministic_2017}, molecules \cite{wang_turning_2019}, or epitaxial QDs \cite{najer_gated_2019}. In this context, recent studies have demonstrated the deterministic coupling of single pQDs to an open Fabry-Pérot microcavity, leading to room temperature single photon emission with significant spectral narrowing due to cavity filtering \cite{farrow_ultranarrow_2023}, but without cQED effects. 

In this work, we demonstrate the deterministic coupling of single pQDs into an open fiber-based Fabry–P\'erot microcavity at cryogenic temperature, achieving a twofold enhancement of the emission rate, corresponding to Purcell factors of up to 3.3 for the smallest cavity mode volume. We then exploit the unique multiplet excitonic fine structure of pQDs by monitoring its spectral evolution under cavity length modulation, showing that the pQD-cavity system provides clear spectral cQED signatures in the weak coupling regime. We model these results by accounting for both static and dynamic effects, including pure dephasing and spectral diffusion, which shape the emitter's spectral properties and impact light-matter coupling in distinct ways. The combination of detailed temporal and spectral investigations of the very same pQD enables reliable determination of cQED parameters and their dependence on the cavity mode volume. Overall, our approach provides a new methodology to estimate the vacuum Rabi coupling $g$ - up to 40~\si{\micro eV} in our case - while offering an original means to delineate the respective contributions of pure dephasing and spectral diffusion to the emitter linewidth.

\section*{Results and discussion}

\subsection*{Coupling of a single perovskite quantum dot with a tunable fiber microcavity}

\begin{figure}[!ht]
\includegraphics[width=1\columnwidth]{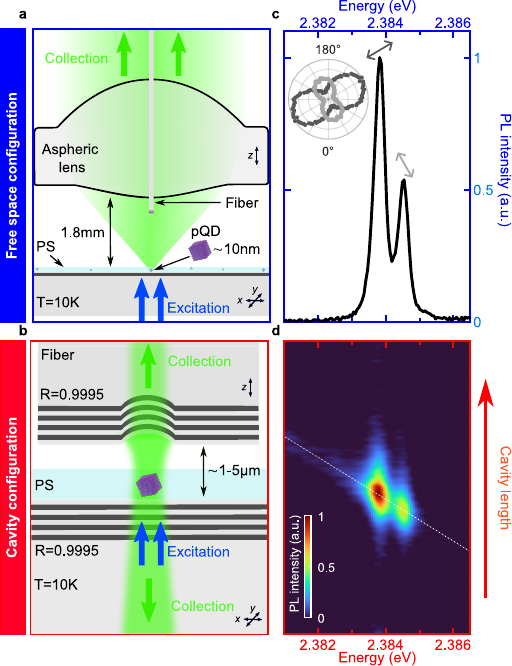}
\caption{\label{fig:setup_scheme} Sketch of the micro-photoluminescence (PL) setup at cryogenic temperature (T=10~\si{\kelvin}) in (a) free space and (b) cavity configurations. PS: Polystyrene, pQD: perovskite Quantum Dot, R: Reflectivity. (c) Free space PL spectrum of a single CsPbBr$_{3}$ pQD (here, pQD1) at 10~\si{\kelvin} under 446~\si{\nano\meter} (2.78~\si{\electronvolt}) cw laser excitation, showing the exciton fine structure with two linearly cross-polarized lines of linewidths 350~\si{\micro\electronvolt} and 290~\si{\micro\electronvolt} for the lower and higher energy peaks, respectively. Inset: Emission polarization diagram of the excitonic doublet where the dark and light gray dots represent the intensity of each line. (d) 2D plot of the emission spectrum of the pQD coupled to the fiber microcavity, when the longitudinal mode $p=14$ of intrinsic linewidth $\hbar\kappa\simeq$ 110~\si{\micro\electronvolt} is tuned across the excitonic lines. The color scale encodes the emission intensity and the white dotted line shows the energy tuning of the longitudinal cavity mode as the cavity length is varied.}
\end{figure}

% SETUP
The synthesis of lead halide CsPbBr$_3$ pQDs is carried out in-house using the hot injection method. Cubic-shaped pQDs with approximate edge sizes of 10~\si{\nano\meter} are obtained in solution. In order to study the optical properties of individual pQDs either in free space or in cavity, the solution is diluted in a polystyrene-toluene mixture and spin-coated on a high-reflectivity planar dielectric mirror (see Methods). The resulting sample then consists of highly dispersed and randomly oriented pQDs embedded in a polystyrene matrix, which prevents the pQDs from degradation with air and humidity and is therefore crucial for the pQDs emission stabilization \cite{raino_underestimated_2019}. Micro-photoluminescence (PL) experiments are performed at cryogenic temperature (10~\si{\kelvin}), using backside excitation through the planar mirror beyond the stop band and collection by a drilled aspheric lens (Figure~\ref{fig:setup_scheme}a). Once a single pQD is characterized in free space, this setup further enables coupling the very same pQD to an optical microcavity (Figure~\ref{fig:setup_scheme}b), where the second mirror is fabricated at the tip of an optical fiber inserted in the drilled lens \cite{jeantetWidelyTunableSinglePhoton2016b,borel_telecom_2023}. Spectral matching is then achieved by precisely adjusting the distance between the mirrors to tune the cavity resonance, while spatial matching is achieved by laterally displacing the planar mirror. Furthermore, the cavity can be opened and closed at will while keeping all other parameters (temperature, excitation, local electrostatic environment) constant, providing a unique opportunity to directly compare the properties of an individual pQD in free space and in cQED regimes (see Methods and Supplemental Information, section~I, for more details on the microcavity implementation and the optical properties of the mirrors).

%TC:ignore
%TC:endignore

% CAVITY COUPLING
Figure~\ref{fig:setup_scheme}c shows the free space PL spectrum of a single pQD. Here, the spectrum consists of a cross-linearly polarized doublet at 2.38~\si{\electronvolt} characterized by linewidths of 350~\si{\micro eV} and 290~\si{\micro eV} for the lower and higher energy peaks respectively, and a splitting $\Delta=700$~\si{\micro eV}. The full pQD spectrum showing the trion and biexciton complexes, along with another example of spectrum where an excitonic triplet can be resolved, are shown in the Supplemental Information (see SI, section~II). We have previously shown that the observation of a doublet, instead of the expected triplet for CsPbBr$_{3}$ pQDs at cryogenic temperatures, can be explained by the orientation of the pQD (and thus of its three emission dipoles) within the polystyrene matrix relatively to the direction of observation \cite{amara_spectral_2023}. For a pQD spectrum where only one doublet is observed, it can be assumed that one of the emission dipoles is along the direction of observation, i.e. here along the microcavity axis (z-axis), while the other two are orthogonal. In the following, we focus on such a pQD, which we model as a pair of orthogonal dipoles, hence optimally coupled to the cavity mode (the effect of the random z positioning of pQDs is discussed in SI, section~III). By varying the cavity length, we observe in Figure~\ref{fig:setup_scheme}d that emission is detected in resonant condition only, when the longitudinal cavity mode of linewidth $\hbar\kappa\simeq 110~\si{\micro eV}$ is tuned at one of the excitonic doublet energies. Note that the mechanical vibrations of the open cavity induce an apparent broadening of the cavity modes with $2\sqrt{2\ln 2}\ \hbar\sigma_{\text{vib}}\simeq1.6~ \si{\milli\electronvolt}$ for the lowest accessible mode. This observation of the pQD emission through the highly reflective mirrors of the cavity is the first strong evidence for the effective coupling of the pQD to the cavity mode.

\subsection*{Lifetime acceleration of a single perovskite quantum dot in the Purcell regime}

\begin{figure}[!ht]
\includegraphics[width=1\columnwidth]{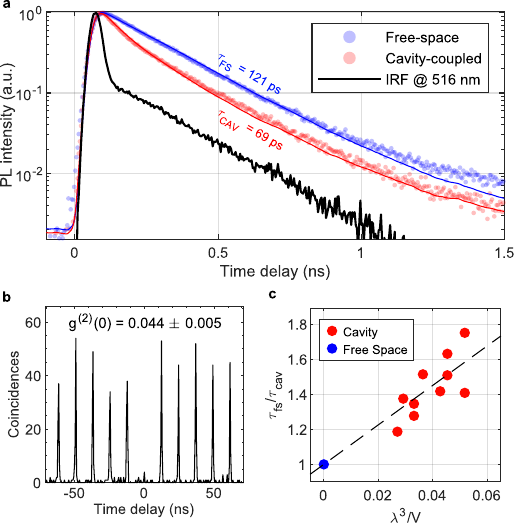}
\caption{\label{fig:TRPL} (a) Time-resolved PL of a single CsPbBr$_3$ pQD (here, pQD2) in free space (blue dots) and coupled to the cavity (red dots). The black line is the Instrumental Response Function (IRF) measured at the pQD emission wavelength. The red and blue solid lines are the fits of the decays using a convolution of the IRF with mono-exponential decays, giving PL lifetimes $\tau_{\text{fs}}=121$~\si{\pico s} in free space and $\tau_{\text{cav}}=69$~\si{\pico s} in cavity, taking into account the photon storage in the cavity. (b) Intensity auto-correlation function measured in free space under pulsed excitation with a Hanbury-Brown and Twiss experiment, where both lines of the excitonic doublet are collected, showing a photon antibunching at zero time delay with $g^{(2)}(0)=0.044\pm0.005$. (c) Acceleration of the emission rate in the cavity, $\tau_{\text{fs}}/\tau_{\text{cav}}$, as a function of the cavity mode volume $V$ in units of $\lambda^3/V$, where $\lambda=520$~\si{\nano\meter} is the average emission wavelength at 10~\si{\kelvin} and $V$ is estimated using finite element simulations.}
\end{figure}

% LIFETIME ACCELERATION
Once a quantum emitter is efficiently coupled to a cavity, an acceleration of its decay rate is expected due to the modification of the electromagnetic density of states. This Purcell effect \cite{Purcell1946} is studied here by analyzing the time-resolved PL of the pQD in both free space and cavity configurations under pulsed excitation. In free space, time-resolved PL can be performed by spectrally integrating the emission of the pQD excitonic doublet, since each dipole emission has identical lifetimes (see SI, section~IV). In cavity, the two emission lines are successively coupled due to the modulation of the pQD-cavity detuning induced by the mechanical vibrations. Therefore, within the measurement duration, the time-resolved PL includes the contribution of both lines. Figure \ref{fig:TRPL}a shows the PL decay of the very same pQD in free space (blue dots) and in cavity (red dots). A shortening of the PL lifetime is observed in the cavity configuration. By convoluting the Instrument Response Function (IRF) with a single exponential decay and accounting for the photon storage in the cavity configuration (see SI, section~IV), PL lifetimes of $\tau_{\text{fs}}=120.8\pm 0.3$~\si{\pico s}  and  $\tau_{\text{cav}}=69.0 \pm 0.3$~\si{\pico s}  are deduced in free space and cavity configurations, respectively. Repeating this procedure yields an acceleration of the emission rate in the cavity by a factor of $\tau_{\text{fs}}/\tau_{\text{cav}}=1.6\pm0.2 $. For all investigated pQDs, an emission acceleration of similar magnitude was systematically observed in the cavity configuration (see SI, Section~IV). We also emphasize that the time-resolved PL of an individual pQD can show significant variations with time (up to 50~\si{\percent} in free space within a few hours), and that only a reconfigurable microcavity, such as the one used here, can provide a reliable acceleration value (see SI, Section~IV where multiple measurements of the pQD PL lifetimes are made successively in free space and in cavity). 

%TC:ignore
%TC:endignore

% PURCELL FACTOR
In a first approach, the pQD is considered as a single two-level system. The photon antibunching observed in the second-order intensity auto-correlation function $g^{(2)}(\tau)$ measured at the excitonic transition (Figure \ref{fig:TRPL}b) indeed shows that the emission from each line of the excitonic doublet does not occur simultaneously. cQED effects in the weak coupling regime are then typically assessed through the Purcell factor $F_p$ characterizing the emission acceleration. The PL decay rate in cavity $\gamma_{\text{cav}}=\tau_{\text{cav}}^{-1}$ is related to the free space PL decay rate $\gamma=\tau_{\text{fs}}^{-1}$ by $\gamma_{\text{cav}}=\gamma(1+\eta_{\text{QY}}F_p)$, where $\eta_{\text{QY}}$ refers to the free space radiative quantum yield of the two-level system transition. The relation between $\gamma$ and $\gamma_{\text{cav}}$ is such that the theoretical expression of $F_p$ matches the usual expression: $F_p=\frac{3}{4\pi^2}\left(\frac{\lambda}{n}\right)^3\frac{Q_{\text{eff}}}{V}$, where $V$ is the cavity mode volume, $n$ the refractive index at the emitter position and $Q_{\text{eff}}$ the effective quality factor which is related to the cavity Q-factor $Q_{\text{cav}}$ and the emitter Q-factor $Q_{\text{em}}$ through $Q_{\text{eff}}^{-1} =Q_{\text{cav}}^{-1}+Q_{\text{em}}^{-1}$ \cite{Auffeves2010}. The quantum yield $\eta_{\text{QY}}$ can be determined experimentally and  is given by the number of photons emitted per excitation pulse at saturation,  $\eta_{\text{QY}}=0.18 \pm 0.02$ for pQD2 (see SI, section~II), allowing us to extract an experimental Purcell factor $F_p=3.3 \pm 1.2$. This value agrees with the theoretical one of $F_p=3.2$ obtained for $Q_{\text{eff}}=2700$, $n=1.5$ and $\lambda^3/V=0.052$. Finally, using our tunable microcavity, we can verify that the PL lifetime acceleration $\tau_{\text{fs}}/\tau_{\text{cav}}$ follows the expected linear trend with the inverse of the cavity mode volume, as shown in Figure \ref{fig:TRPL}c where the cavity length is varied by steps of half a wavelength.

\subsection*{Vacuum Rabi coupling strength}

% CONTEXT
We now take into account the specificities of the hybrid pQD-cavity system in a more comprehensive model, which allows us to assess the vacuum Rabi coupling $g$, a fundamental cQED figure of merit encoding the strength of light-matter coupling. In the case of a simple two-level system, $g$ is related to the acceleration in cavity by $4g^2=(\gamma_{\text{cav}, \delta=0} -\gamma)(\gamma+\gamma^*+\kappa)$, where $\gamma_{\text{cav}, \delta=0}=\tau_{\text{cav}, \delta=0}^{-1}$ is the emitter decay rate in cavity at resonance, $\gamma=\tau_{\text{fs}}^{-1}$ the free space emitter decay rate, $\gamma^*$ the emitter pure dephasing rate and $\kappa$ the cavity loss rate \cite{Auffeves2010}. For most solid-state quantum emitters, the evaluation of $\gamma^*$ is complicated by the presence of spectral diffusion, which is experimentally difficult to distinguish from pure dephasing when it occurs on a time scale shorter than the integration time. In fact, even if the instantaneous PL linewidth is given by $\hbar(\gamma+\gamma^*)$, a typical Gaussian broadening $\sigma_{\text{SD}}$ of the emission line can be caused by spectral diffusion on any time scale between the emitter lifetime and the integration time. Consequently, the measured linewidth is the sum of both contributions and hence only an upper bound for the pure dephasing rate can be extracted from the spectra. Spectral diffusion leads to a time-varying cavity-emitter detuning $\delta$ and the apparent decay rate measured in cavity $\gamma_{\text{cav}}$ is thus smaller than the value expected strictly at resonance $\gamma_{\text{cav}, \delta=0}$. This effect makes it difficult to deduce a reliable value of $g$ from the analysis of the temporal dynamics. 

To obtain an independent estimate of $g$ and determine the transition between the weak and strong light-matter coupling regimes, one can instead analyse the spectral fingerprints of the cavity-emitter system as a function of the emitter-cavity detuning. For very low coupling strengths, the hybrid system emission spectrum is simply the product of the emitter and cavity transmission spectra. When the cavity mode linewidth is smaller than that of the emitter, the cavity acts as a narrow spectral filter. As the coupling $g$ increases, spectral mixing between the emitter and cavity contributions occurs, resulting in more complex spectral signatures which depend on the emitter-cavity detuning. 

\begin{figure*}[ht!]  % figure* pour full column... 
\includegraphics[width=1\textwidth]{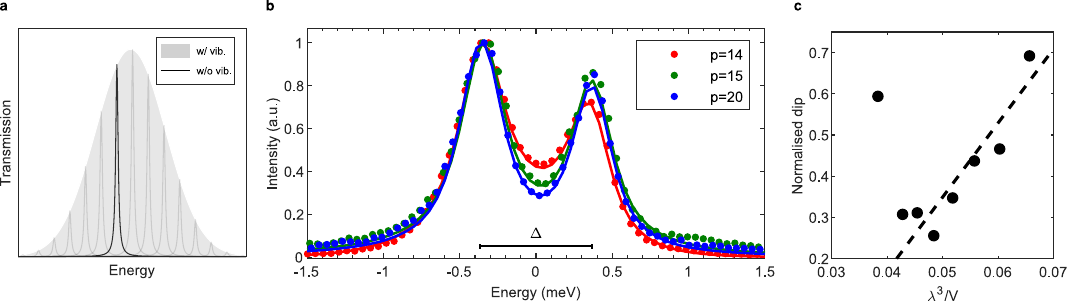}
\caption{\label{fig:envelopeNEW}(a) Scheme of the cavity mode of instantaneous linewidth $\hbar\kappa=110~\si{\micro eV}$ (solid black line) modulated by the mechanical vibrations of the fiber microcavity. At the order $p=14$ for which the fiber and planar mirrors are in contact, the modulation covers a spectral range of $2\sqrt{2\ln 2}\ \hbar\sigma_{\text{vib}}\simeq14\hbar\kappa=1.6~\si{\milli eV}$ (light gray area). (b) Normalized spectral envelopes (dots) of the emission of the coupled pQD-cavity system (here with pQD1), for three longitudinal cavity modes ($p=14$, 15 and 20), along with their fits (solid line, see SI, section V). The doublet splitting is indicated with the $\Delta$ scale. (c) Normalized dip of the spectral envelope, including a simple correction taking into account the cavity vibrations, as a function of the inverse cavity mode volume  $\lambda^3/V$. The dashed line is a guide for the eye.} 
\end{figure*}

\begin{figure}[!ht]
\includegraphics[width=1\columnwidth]{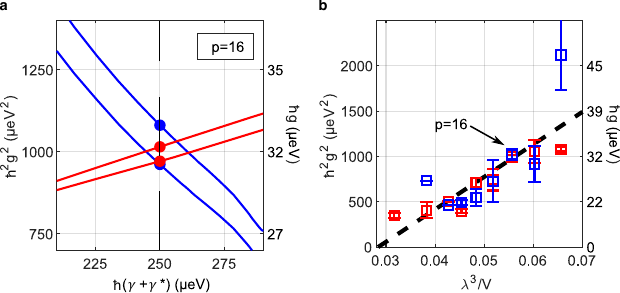}
\caption{\label{fig:rabi_couplingNEW}  (a) Vacuum Rabi coupling strength $g$ as a function of the pQD instantaneous linewidth $\hbar(\gamma+\gamma^*)$. The blue and red lines are obtained from the analysis of the spectral envelope and PL decay, respectively, measured for the same pQD (here, pQD1) coupled to the cavity mode $p=16$. The spectral envelope and PL decay analysis were done on two sets of data, leading to two pairs of independent curves. The dots are the intersection of these curves with $\hbar(\gamma+\gamma^*)=250$~\si{\micro eV}. (b) Vacuum Rabi coupling strength $g$ as a function of the cavity mode volume $V$ in units of $\lambda^3/V$. The empty squares are the mean intersection values corresponding to the crossing dots for $\hbar(\gamma+\gamma^*)=250\pm50$~\si{\micro eV}. The error bars are deduced from the two sets of data. The dashed line is a guide for the eye.}
\end{figure}

% SPECTRAL ENVELOPE
With our flexible platform, the detuning can be modulated either externally or simply by exploiting the spontaneous cavity's mechanical vibrations. In the latter case, the cavity length is indeed modulated at sub-kHz frequencies, leading to an apparent cavity linewidth $2\sqrt{2\ln 2}\ \hbar\sigma_{\text{vib}}\simeq8~\si{\milli\electronvolt}\simeq 70 \hbar\kappa$ for most accessible longitudinal modes. These vibrations can be reduced by a mechanical contact of the two mirrors, i.e. for the lowest accessible mode $p=14$ where $2\sqrt{2\ln 2}\ \hbar\sigma_{\text{vib}}=1.6~\si{\milli\electronvolt}$. Similarly, while the coupled pQD-cavity system is characterized by an instantaneous Lorentzian spectrum for each cavity length, the emission spectrum of the modulated pQD-cavity system shows a spectral envelope that is displayed in Figure~\ref{fig:envelopeNEW}a for three longitudinal cavity modes ($p=14$, 15, and 20). For each mode, the envelope features two main peaks resulting from the spectral matching of the cavity with each line of the excitonic doublet. The shape of the envelope, particularly the central minimum hereafter called \textit{dip}, is strongly influenced by the emitter-cavity coupling. Figure \ref{fig:envelopeNEW}c summarizes the measured normalized dip depths as a function of the inverse cavity mode volume $\lambda^3/V$ (see Methods for the dip evaluation). 
The observed linear trend of the normalized dip reflects the modification of the light matter coupling when varying the volume \footnote{We note that the simple extraction of the dip for $\lambda^3/V\simeq 0.038$ is far from the general trend. However, this discrepancy is less critical when considering the full theoretical analysis as shown in Figure~\ref{fig:rabi_couplingNEW}b.}.
Hence, this feature can be used to assess $g$.

%TC:ignore

%TC:endignore

% MODEL AND DATA ANALYSIS
In the case of a single two-level system coupled to a cavity mode, the spectral envelope and its dependence with $g$ can be calculated analytically (see SI, section~V). Here, we describe the two non-degenerate and orthogonally polarized emission lines (Figure \ref{fig:setup_scheme}c) as two independent and mutually exclusive two-level systems coupled to distinct cavity modes, resulting in two independent subsystems. This framework is indeed valid because the two cavity polarization modes are degenerate and the polarization basis can be chosen arbitrarily aligned with that of the pQD. Therefore, the two subsystems can be incoherently summed in the model. 

Moreover, the resulting spectral envelope not only depends on the vacuum Rabi coupling $g$, but also on the spectral diffusion amplitude $\sigma_{\text{SD}}$ and the instantaneous linewidth $\hbar (\gamma+\gamma^*)$, all the other parameters being experimentally determined. Since the total free space PL linewidth is the sum of the pure dephasing and spectral diffusion contributions, a relationship between $\sigma_{\text{SD}}$ and $(\gamma+\gamma^*)$ can be established (see SI, section~V), reducing the model to two free parameters. Similarly, the PL decays that were previously analyzed in a simplified picture are now modeled including the varying detuning $\delta$ induced by the cavity vibrations and the spectral diffusion (see SI, section~V). Our model shows that higher pure dephasing requires, firstly, higher $g$ values to account for the experimental PL decay rate and, secondly, lower $g$ values to fit the experimental spectral envelope profile. Hence, the simultaneous fitting of both experimental inputs (PL decay and spectral envelope) yields a unique intersection from which an accurate estimate of $g$ can be deduced, as shown in Figure \ref{fig:rabi_couplingNEW}a for the mode $p=16$. This opposite trend can be understood by considering the dominant dimensionless parameters governing the PL decay and the spectral envelope, which are given by $\frac{g^2}{(\kappa+\gamma^*)\gamma}$ and $\frac{\gamma^*}{\kappa+\gamma^*}\frac{g^2}{(\kappa+\gamma^*)\gamma}$, respectively. To keep them constant while increasing the instantaneous linewidth ($\simeq \gamma^*$), an increase in $g$ is required for the former, whereas a decrease is needed for the latter. Hence, the intersection of the two curves yields all together $g$, $\hbar(\gamma+\gamma^*)$ and $\sigma_{\text{SD}}$. By reproducing this analysis for all the accessible mode volumes in our platform, the same behavior is obtained with a consistent crossing at $\hbar(\gamma+\gamma^*)=250\pm50$~\si{\micro eV} (see SI, section~V). Figure~\ref{fig:rabi_couplingNEW}b presents the corresponding values of $g$ as a function of the inverse mode volume and shows the expected increase of the Rabi coupling, $g^2$ following a linear trend with $\lambda^3/V$. We stress that this comprehensive analysis, based on two complementary experiments, provides a reliable value not only for the vacuum Rabi coupling strength $g$ but also for the emitter's instantaneous linewidth, which are key parameters for quantum technology applications. 

%TC:ignore
%TC:endignore

In conclusion, we have demonstrated the efficient coupling of single pQDs into an open fiber Fabry-Pérot microcavity, which offers great flexibility in spatial and spectral tuning. The deterministic and reversible coupling of fully characterized single photon emitters, while maintaining the same excitation and local environment, enables insightful cQED experiments in the weak light-matter coupling regime. We systematically observe an acceleration of the emission rate with a maximum twofold acceleration. In a simple two-level system picture, where the radiative quantum yield of the pQD could be estimated to be on the order of $20$~\si{\percent}, we deduce that this acceleration corresponds to a Purcell factor of $3.3 \pm 1.2$. Beyond this first order approach, we developed a comprehensive theoretical model that encompasses advanced spectral properties of the pQDs, i.e. the excitonic fine structure and the linewidth broadening due to pure dephasing and spectral diffusion, in order to accurately analyse the temporal and spectral signatures of the hybrid pQD-cavity system. Using this theoretical framework to analyse two independent sets of experimental results, we extracted key metrics of our pQD-cavity platform: the instantaneous pQD linewidth ($\hbar(\gamma+\gamma^*)=250\pm50$~\si{\micro \electronvolt}, free from any spectral diffusion contribution) and the light-matter vacuum Rabi coupling strength ($g=40\pm 10$~\si{\micro \electronvolt} for the smallest mode volume). On the one hand, these results hold great promise for the emission of cavity-enhanced indistinguishable photons. Indeed, according to the work of T. Grange et al. \cite{grange_cavity-funneled_2015}, the indistinguishability of photons emitted through the cavity can be estimated to be up to 16\%, which is a fourfold improvement over the free space value. This value is obtained while maintaining a high single-photon efficiency of $\sim60\%$. On the other hand, this study also demonstrates that achieving the strong coupling regime is within reach with this platform using for instance pQDs with optimized ligand stabilization, for which linewidths as narrow as $30$~\si{\micro \electronvolt} have been reported \cite{utzat_coherent_2019}, and by further reducing the cavity volume as we have already done with other emitters \cite{borel_telecom_2023} while enhancing the cavity finesse to a reasonable extent \cite{wang_turning_2019}.

%TC:ignore
\bibliography{biblio_article_v3}% Produces the bibliography via BibTeX.

\begin{thebibliography}{30}%
\makeatletter
\providecommand \@ifxundefined [1]{%
 \@ifx{#1\undefined}
}%
\providecommand \@ifnum [1]{%
 \ifnum #1\expandafter \@firstoftwo
 \else \expandafter \@secondoftwo
 \fi
}%
\providecommand \@ifx [1]{%
 \ifx #1\expandafter \@firstoftwo
 \else \expandafter \@secondoftwo
 \fi
}%
\providecommand \natexlab [1]{#1}%
\providecommand \enquote  [1]{``#1''}%
\providecommand \bibnamefont  [1]{#1}%
\providecommand \bibfnamefont [1]{#1}%
\providecommand \citenamefont [1]{#1}%
\providecommand \href@noop [0]{\@secondoftwo}%
\providecommand \href [0]{\begingroup \@sanitize@url \@href}%
\providecommand \@href[1]{\@@startlink{#1}\@@href}%
\providecommand \@@href[1]{\endgroup#1\@@endlink}%
\providecommand \@sanitize@url [0]{\catcode `\\12\catcode `\$12\catcode
  `\&12\catcode `\#12\catcode `\^12\catcode `\_12\catcode `\%12\relax}%
\providecommand \@@startlink[1]{}%
\providecommand \@@endlink[0]{}%
\providecommand \url  [0]{\begingroup\@sanitize@url \@url }%
\providecommand \@url [1]{\endgroup\@href {#1}{\urlprefix }}%
\providecommand \urlprefix  [0]{URL }%
\providecommand \Eprint [0]{\href }%
\providecommand \doibase [0]{https://doi.org/}%
\providecommand \selectlanguage [0]{\@gobble}%
\providecommand \bibinfo  [0]{\@secondoftwo}%
\providecommand \bibfield  [0]{\@secondoftwo}%
\providecommand \translation [1]{[#1]}%
\providecommand \BibitemOpen [0]{}%
\providecommand \bibitemStop [0]{}%
\providecommand \bibitemNoStop [0]{.\EOS\space}%
\providecommand \EOS [0]{\spacefactor3000\relax}%
\providecommand \BibitemShut  [1]{\csname bibitem#1\endcsname}%
\let\auto@bib@innerbib\@empty
%</preamble>
\bibitem [{\citenamefont {Protesescu}\ \emph {et~al.}(2015)\citenamefont
  {Protesescu}, \citenamefont {Yakunin}, \citenamefont {Bodnarchuk},
  \citenamefont {Krieg}, \citenamefont {Caputo}, \citenamefont {Hendon},
  \citenamefont {Yang}, \citenamefont {Walsh},\ and\ \citenamefont
  {Kovalenko}}]{Protesescu}%
  \BibitemOpen
  \bibfield  {author} {\bibinfo {author} {\bibfnamefont {L.}~\bibnamefont
  {Protesescu}}, \bibinfo {author} {\bibfnamefont {S.}~\bibnamefont {Yakunin}},
  \bibinfo {author} {\bibfnamefont {M.~I.}\ \bibnamefont {Bodnarchuk}},
  \bibinfo {author} {\bibfnamefont {F.}~\bibnamefont {Krieg}}, \bibinfo
  {author} {\bibfnamefont {R.}~\bibnamefont {Caputo}}, \bibinfo {author}
  {\bibfnamefont {C.~H.}\ \bibnamefont {Hendon}}, \bibinfo {author}
  {\bibfnamefont {R.~X.}\ \bibnamefont {Yang}}, \bibinfo {author}
  {\bibfnamefont {A.}~\bibnamefont {Walsh}},\ and\ \bibinfo {author}
  {\bibfnamefont {M.~V.}\ \bibnamefont {Kovalenko}},\ }\bibfield  {title}
  {\bibinfo {title} {Nanocrystals of cesium lead halide perovskites
  ({CsPbX}$_{\textrm{3}}$, {X = Cl}, {Br}, and {I}): Novel optoelectronic
  materials showing bright emission with wide color gamut},\ }\href
  {https://doi.org/10.1021/nl5048779} {\bibfield  {journal} {\bibinfo
  {journal} {Nano Letters}\ }\textbf {\bibinfo {volume} {15}},\ \bibinfo
  {pages} {3692} (\bibinfo {year} {2015})}\BibitemShut {NoStop}%
\bibitem [{\citenamefont {Sutherland}\ and\ \citenamefont
  {Sargent}(2016)}]{sutherland_perovskite_2016}%
  \BibitemOpen
  \bibfield  {author} {\bibinfo {author} {\bibfnamefont {B.~R.}\ \bibnamefont
  {Sutherland}}\ and\ \bibinfo {author} {\bibfnamefont {E.~H.}\ \bibnamefont
  {Sargent}},\ }\bibfield  {title} {\bibinfo {title} {Perovskite photonic
  sources},\ }\href {https://doi.org/10.1038/nphoton.2016.62} {\bibfield
  {journal} {\bibinfo  {journal} {Nature Photonics}\ }\textbf {\bibinfo
  {volume} {10}},\ \bibinfo {pages} {295} (\bibinfo {year} {2016})}\BibitemShut
  {NoStop}%
\bibitem [{\citenamefont {Aharonovich}\ \emph {et~al.}(2016)\citenamefont
  {Aharonovich}, \citenamefont {Englund},\ and\ \citenamefont
  {Toth}}]{aharonovich_solid-state_2016}%
  \BibitemOpen
  \bibfield  {author} {\bibinfo {author} {\bibfnamefont {I.}~\bibnamefont
  {Aharonovich}}, \bibinfo {author} {\bibfnamefont {D.}~\bibnamefont
  {Englund}},\ and\ \bibinfo {author} {\bibfnamefont {M.}~\bibnamefont
  {Toth}},\ }\bibfield  {title} {\bibinfo {title} {Solid-state single-photon
  emitters},\ }\href {https://doi.org/10.1038/nphoton.2016.186} {\bibfield
  {journal} {\bibinfo  {journal} {Nature Photonics}\ }\textbf {\bibinfo
  {volume} {10}},\ \bibinfo {pages} {631} (\bibinfo {year} {2016})}\BibitemShut
  {NoStop}%
\bibitem [{\citenamefont {Hu}\ \emph {et~al.}(2015)\citenamefont {Hu},
  \citenamefont {Zhang}, \citenamefont {Sun}, \citenamefont {Yin},
  \citenamefont {Lv}, \citenamefont {Zhang}, \citenamefont {Yu}, \citenamefont
  {Wang}, \citenamefont {Zhang},\ and\ \citenamefont
  {Xiao}}]{hu_superior_2015}%
  \BibitemOpen
  \bibfield  {author} {\bibinfo {author} {\bibfnamefont {F.}~\bibnamefont
  {Hu}}, \bibinfo {author} {\bibfnamefont {H.}~\bibnamefont {Zhang}}, \bibinfo
  {author} {\bibfnamefont {C.}~\bibnamefont {Sun}}, \bibinfo {author}
  {\bibfnamefont {C.}~\bibnamefont {Yin}}, \bibinfo {author} {\bibfnamefont
  {B.}~\bibnamefont {Lv}}, \bibinfo {author} {\bibfnamefont {C.}~\bibnamefont
  {Zhang}}, \bibinfo {author} {\bibfnamefont {W.~W.}\ \bibnamefont {Yu}},
  \bibinfo {author} {\bibfnamefont {X.}~\bibnamefont {Wang}}, \bibinfo {author}
  {\bibfnamefont {Y.}~\bibnamefont {Zhang}},\ and\ \bibinfo {author}
  {\bibfnamefont {M.}~\bibnamefont {Xiao}},\ }\bibfield  {title} {\bibinfo
  {title} {Superior {Optical} {Properties} of {Perovskite} {Nanocrystals} as
  {Single} {Photon} {Emitters}},\ }\href
  {https://doi.org/10.1021/acsnano.5b05769} {\bibfield  {journal} {\bibinfo
  {journal} {ACS Nano}\ }\textbf {\bibinfo {volume} {9}},\ \bibinfo {pages}
  {12410} (\bibinfo {year} {2015})}\BibitemShut {NoStop}%
\bibitem [{\citenamefont {Park}\ \emph {et~al.}(2015)\citenamefont {Park},
  \citenamefont {Guo}, \citenamefont {Makarov},\ and\ \citenamefont
  {Klimov}}]{park_room_2015}%
  \BibitemOpen
  \bibfield  {author} {\bibinfo {author} {\bibfnamefont {Y.-S.}\ \bibnamefont
  {Park}}, \bibinfo {author} {\bibfnamefont {S.}~\bibnamefont {Guo}}, \bibinfo
  {author} {\bibfnamefont {N.~S.}\ \bibnamefont {Makarov}},\ and\ \bibinfo
  {author} {\bibfnamefont {V.~I.}\ \bibnamefont {Klimov}},\ }\bibfield  {title}
  {\bibinfo {title} {Room {Temperature} {Single}-{Photon} {Emission} from
  {Individual} {Perovskite} {Quantum} {Dots}},\ }\href
  {https://doi.org/10.1021/acsnano.5b04584} {\bibfield  {journal} {\bibinfo
  {journal} {ACS Nano}\ }\textbf {\bibinfo {volume} {9}},\ \bibinfo {pages}
  {10386} (\bibinfo {year} {2015})}\BibitemShut {NoStop}%
\bibitem [{\citenamefont {Zhu}\ \emph {et~al.}(2022)\citenamefont {Zhu},
  \citenamefont {Marczak}, \citenamefont {Feld}, \citenamefont {Boehme},
  \citenamefont {Bernasconi}, \citenamefont {Moskalenko}, \citenamefont
  {Cherniukh}, \citenamefont {Dirin}, \citenamefont {Bodnarchuk}, \citenamefont
  {Kovalenko},\ and\ \citenamefont {Rainò}}]{zhu_room-temperature_2022}%
  \BibitemOpen
  \bibfield  {author} {\bibinfo {author} {\bibfnamefont {C.}~\bibnamefont
  {Zhu}}, \bibinfo {author} {\bibfnamefont {M.}~\bibnamefont {Marczak}},
  \bibinfo {author} {\bibfnamefont {L.}~\bibnamefont {Feld}}, \bibinfo {author}
  {\bibfnamefont {S.~C.}\ \bibnamefont {Boehme}}, \bibinfo {author}
  {\bibfnamefont {C.}~\bibnamefont {Bernasconi}}, \bibinfo {author}
  {\bibfnamefont {A.}~\bibnamefont {Moskalenko}}, \bibinfo {author}
  {\bibfnamefont {I.}~\bibnamefont {Cherniukh}}, \bibinfo {author}
  {\bibfnamefont {D.}~\bibnamefont {Dirin}}, \bibinfo {author} {\bibfnamefont
  {M.~I.}\ \bibnamefont {Bodnarchuk}}, \bibinfo {author} {\bibfnamefont
  {M.~V.}\ \bibnamefont {Kovalenko}},\ and\ \bibinfo {author} {\bibfnamefont
  {G.}~\bibnamefont {Rainò}},\ }\bibfield  {title} {\bibinfo {title}
  {Room-temperature, highly pure single-photon sources from all-inorganic lead
  halide perovskite quantum dots},\ }\href
  {https://doi.org/10.1021/acs.nanolett.2c00756} {\bibfield  {journal}
  {\bibinfo  {journal} {Nano Letters}\ }\textbf {\bibinfo {volume} {22}},\
  \bibinfo {pages} {3751} (\bibinfo {year} {2022})}\BibitemShut {NoStop}%
\bibitem [{\citenamefont {Utzat}\ \emph {et~al.}(2019)\citenamefont {Utzat},
  \citenamefont {Sun}, \citenamefont {Kaplan}, \citenamefont {Krieg},
  \citenamefont {Ginterseder}, \citenamefont {Spokoyny}, \citenamefont {Klein},
  \citenamefont {Shulenberger}, \citenamefont {Perkinson}, \citenamefont
  {Kovalenko},\ and\ \citenamefont {Bawendi}}]{utzat_coherent_2019}%
  \BibitemOpen
  \bibfield  {author} {\bibinfo {author} {\bibfnamefont {H.}~\bibnamefont
  {Utzat}}, \bibinfo {author} {\bibfnamefont {W.}~\bibnamefont {Sun}}, \bibinfo
  {author} {\bibfnamefont {A.~E.~K.}\ \bibnamefont {Kaplan}}, \bibinfo {author}
  {\bibfnamefont {F.}~\bibnamefont {Krieg}}, \bibinfo {author} {\bibfnamefont
  {M.}~\bibnamefont {Ginterseder}}, \bibinfo {author} {\bibfnamefont
  {B.}~\bibnamefont {Spokoyny}}, \bibinfo {author} {\bibfnamefont {N.~D.}\
  \bibnamefont {Klein}}, \bibinfo {author} {\bibfnamefont {K.~E.}\ \bibnamefont
  {Shulenberger}}, \bibinfo {author} {\bibfnamefont {C.~F.}\ \bibnamefont
  {Perkinson}}, \bibinfo {author} {\bibfnamefont {M.~V.}\ \bibnamefont
  {Kovalenko}},\ and\ \bibinfo {author} {\bibfnamefont {M.~G.}\ \bibnamefont
  {Bawendi}},\ }\bibfield  {title} {\bibinfo {title} {Coherent single-photon
  emission from colloidal lead halide perovskite quantum dots},\ }\href
  {https://doi.org/10.1126/science.aau7392} {\bibfield  {journal} {\bibinfo
  {journal} {Science}\ }\textbf {\bibinfo {volume} {363}},\ \bibinfo {pages}
  {1068} (\bibinfo {year} {2019})}\BibitemShut {NoStop}%
\bibitem [{\citenamefont {Lv}\ \emph {et~al.}(2019)\citenamefont {Lv},
  \citenamefont {Yin}, \citenamefont {Zhang}, \citenamefont {Yu}, \citenamefont
  {Wang}, \citenamefont {Zhang},\ and\ \citenamefont {Xiao}}]{lv_quantum_2019}%
  \BibitemOpen
  \bibfield  {author} {\bibinfo {author} {\bibfnamefont {Y.}~\bibnamefont
  {Lv}}, \bibinfo {author} {\bibfnamefont {C.}~\bibnamefont {Yin}}, \bibinfo
  {author} {\bibfnamefont {C.}~\bibnamefont {Zhang}}, \bibinfo {author}
  {\bibfnamefont {W.~W.}\ \bibnamefont {Yu}}, \bibinfo {author} {\bibfnamefont
  {X.}~\bibnamefont {Wang}}, \bibinfo {author} {\bibfnamefont {Y.}~\bibnamefont
  {Zhang}},\ and\ \bibinfo {author} {\bibfnamefont {M.}~\bibnamefont {Xiao}},\
  }\bibfield  {title} {\bibinfo {title} {Quantum {Interference} in a {Single}
  {Perovskite} {Nanocrystal}},\ }\href
  {https://doi.org/10.1021/acs.nanolett.9b01237} {\bibfield  {journal}
  {\bibinfo  {journal} {Nano Letters}\ }\textbf {\bibinfo {volume} {19}},\
  \bibinfo {pages} {4442} (\bibinfo {year} {2019})}\BibitemShut {NoStop}%
\bibitem [{\citenamefont {Lv}\ \emph {et~al.}(2021)\citenamefont {Lv},
  \citenamefont {Yin}, \citenamefont {Zhang}, \citenamefont {Wang},
  \citenamefont {Yu},\ and\ \citenamefont {Xiao}}]{lv_exciton-acoustic_2021}%
  \BibitemOpen
  \bibfield  {author} {\bibinfo {author} {\bibfnamefont {Y.}~\bibnamefont
  {Lv}}, \bibinfo {author} {\bibfnamefont {C.}~\bibnamefont {Yin}}, \bibinfo
  {author} {\bibfnamefont {C.}~\bibnamefont {Zhang}}, \bibinfo {author}
  {\bibfnamefont {X.}~\bibnamefont {Wang}}, \bibinfo {author} {\bibfnamefont
  {Z.-G.}\ \bibnamefont {Yu}},\ and\ \bibinfo {author} {\bibfnamefont
  {M.}~\bibnamefont {Xiao}},\ }\bibfield  {title} {\bibinfo {title}
  {Exciton-acoustic phonon coupling revealed by resonant excitation of single
  perovskite nanocrystals},\ }\href
  {https://doi.org/10.1038/s41467-021-22486-5} {\bibfield  {journal} {\bibinfo
  {journal} {Nature Communications}\ }\textbf {\bibinfo {volume} {12}},\
  \bibinfo {pages} {2192} (\bibinfo {year} {2021})}\BibitemShut {NoStop}%
\bibitem [{\citenamefont {Kaplan}\ \emph {et~al.}(2023)\citenamefont {Kaplan},
  \citenamefont {Krajewska}, \citenamefont {Proppe}, \citenamefont {Sun},
  \citenamefont {Sverko}, \citenamefont {Berkinsky}, \citenamefont {Utzat},\
  and\ \citenamefont {Bawendi}}]{kaplan_hongoumandel_2023}%
  \BibitemOpen
  \bibfield  {author} {\bibinfo {author} {\bibfnamefont {A.~E.~K.}\
  \bibnamefont {Kaplan}}, \bibinfo {author} {\bibfnamefont {C.~J.}\
  \bibnamefont {Krajewska}}, \bibinfo {author} {\bibfnamefont {A.~H.}\
  \bibnamefont {Proppe}}, \bibinfo {author} {\bibfnamefont {W.}~\bibnamefont
  {Sun}}, \bibinfo {author} {\bibfnamefont {T.}~\bibnamefont {Sverko}},
  \bibinfo {author} {\bibfnamefont {D.~B.}\ \bibnamefont {Berkinsky}}, \bibinfo
  {author} {\bibfnamefont {H.}~\bibnamefont {Utzat}},\ and\ \bibinfo {author}
  {\bibfnamefont {M.~G.}\ \bibnamefont {Bawendi}},\ }\bibfield  {title}
  {\bibinfo {title} {Hong-{Ou}-{Mandel} interference in colloidal
  {CsPbBr}$_{\textrm{3}}$ perovskite nanocrystals},\ }\href
  {https://doi.org/10.1038/s41566-023-01225-w} {\bibfield  {journal} {\bibinfo
  {journal} {Nature Photonics}\ }\textbf {\bibinfo {volume} {17}},\ \bibinfo
  {pages} {775} (\bibinfo {year} {2023})}\BibitemShut {NoStop}%
\bibitem [{\citenamefont {Tamarat}\ \emph {et~al.}(2020)\citenamefont
  {Tamarat}, \citenamefont {Hou}, \citenamefont {Trebbia}, \citenamefont
  {Swarnkar}, \citenamefont {Biadala}, \citenamefont {Louyer}, \citenamefont
  {Bodnarchuk}, \citenamefont {Kovalenko}, \citenamefont {Even},\ and\
  \citenamefont {Lounis}}]{tamarat_dark_2020}%
  \BibitemOpen
  \bibfield  {author} {\bibinfo {author} {\bibfnamefont {P.}~\bibnamefont
  {Tamarat}}, \bibinfo {author} {\bibfnamefont {L.}~\bibnamefont {Hou}},
  \bibinfo {author} {\bibfnamefont {J.-B.}\ \bibnamefont {Trebbia}}, \bibinfo
  {author} {\bibfnamefont {A.}~\bibnamefont {Swarnkar}}, \bibinfo {author}
  {\bibfnamefont {L.}~\bibnamefont {Biadala}}, \bibinfo {author} {\bibfnamefont
  {Y.}~\bibnamefont {Louyer}}, \bibinfo {author} {\bibfnamefont {M.~I.}\
  \bibnamefont {Bodnarchuk}}, \bibinfo {author} {\bibfnamefont {M.~V.}\
  \bibnamefont {Kovalenko}}, \bibinfo {author} {\bibfnamefont {J.}~\bibnamefont
  {Even}},\ and\ \bibinfo {author} {\bibfnamefont {B.}~\bibnamefont {Lounis}},\
  }\bibfield  {title} {\bibinfo {title} {The dark exciton ground state promotes
  photon-pair emission in individual perovskite nanocrystals},\ }\href
  {https://doi.org/10.1038/s41467-020-19740-7} {\bibfield  {journal} {\bibinfo
  {journal} {Nature Communications}\ }\textbf {\bibinfo {volume} {11}},\
  \bibinfo {pages} {6001} (\bibinfo {year} {2020})}\BibitemShut {NoStop}%
\bibitem [{\citenamefont {Wang}\ \emph {et~al.}(2021)\citenamefont {Wang},
  \citenamefont {Rasmita}, \citenamefont {Long}, \citenamefont {Chen},
  \citenamefont {Zhang}, \citenamefont {Garcia}, \citenamefont {Cai},
  \citenamefont {Xiong},\ and\ \citenamefont {Gao}}]{WangOptically}%
  \BibitemOpen
  \bibfield  {author} {\bibinfo {author} {\bibfnamefont {Z.}~\bibnamefont
  {Wang}}, \bibinfo {author} {\bibfnamefont {A.}~\bibnamefont {Rasmita}},
  \bibinfo {author} {\bibfnamefont {G.}~\bibnamefont {Long}}, \bibinfo {author}
  {\bibfnamefont {D.}~\bibnamefont {Chen}}, \bibinfo {author} {\bibfnamefont
  {C.}~\bibnamefont {Zhang}}, \bibinfo {author} {\bibfnamefont {O.~G.}\
  \bibnamefont {Garcia}}, \bibinfo {author} {\bibfnamefont {H.}~\bibnamefont
  {Cai}}, \bibinfo {author} {\bibfnamefont {Q.}~\bibnamefont {Xiong}},\ and\
  \bibinfo {author} {\bibfnamefont {W.-b.}\ \bibnamefont {Gao}},\ }\bibfield
  {title} {\bibinfo {title} {Optically driven giant superbunching from a single
  perovskite quantum dot},\ }\href
  {https://doi.org/https://doi.org/10.1002/adom.202100879} {\bibfield
  {journal} {\bibinfo  {journal} {Advanced Optical Materials}\ }\textbf
  {\bibinfo {volume} {9}},\ \bibinfo {pages} {2100879} (\bibinfo {year}
  {2021})}\BibitemShut {NoStop}%
\bibitem [{\citenamefont {Rain{\`o}}\ \emph {et~al.}(2018)\citenamefont
  {Rain{\`o}}, \citenamefont {Becker}, \citenamefont {Bodnarchuk},
  \citenamefont {Mahrt}, \citenamefont {Kovalenko},\ and\ \citenamefont
  {St{\"o}ferle}}]{raino_superfluorescence_2018}%
  \BibitemOpen
  \bibfield  {author} {\bibinfo {author} {\bibfnamefont {G.}~\bibnamefont
  {Rain{\`o}}}, \bibinfo {author} {\bibfnamefont {M.~A.}\ \bibnamefont
  {Becker}}, \bibinfo {author} {\bibfnamefont {M.~I.}\ \bibnamefont
  {Bodnarchuk}}, \bibinfo {author} {\bibfnamefont {R.~F.}\ \bibnamefont
  {Mahrt}}, \bibinfo {author} {\bibfnamefont {M.~V.}\ \bibnamefont
  {Kovalenko}},\ and\ \bibinfo {author} {\bibfnamefont {T.}~\bibnamefont
  {St{\"o}ferle}},\ }\bibfield  {title} {\bibinfo {title} {Superfluorescence
  from lead halide perovskite quantum dot superlattices},\ }\href
  {https://doi.org/10.1038/s41586-018-0683-0} {\bibfield  {journal} {\bibinfo
  {journal} {Nature}\ }\textbf {\bibinfo {volume} {563}},\ \bibinfo {pages}
  {671} (\bibinfo {year} {2018})}\BibitemShut {NoStop}%
\bibitem [{\citenamefont {Purcell}(1946)}]{Purcell1946}%
  \BibitemOpen
  \bibfield  {author} {\bibinfo {author} {\bibfnamefont {E.~M.}\ \bibnamefont
  {Purcell}},\ }\bibfield  {title} {\bibinfo {title} {Spontaneous emission
  probabilities at radio frequencies.},\ }\bibfield  {journal} {\bibinfo
  {journal} {Phys. Rev.}\ }\textbf {\bibinfo {volume} {69}},\ \href
  {https://doi.org/10.1103/PhysRev.69.674.2} {10.1103/PhysRev.69.674.2}
  (\bibinfo {year} {1946})\BibitemShut {NoStop}%
\bibitem [{\citenamefont {Birnbaum}\ \emph {et~al.}(2005)\citenamefont
  {Birnbaum}, \citenamefont {Boca}, \citenamefont {Miller}, \citenamefont
  {Boozer}, \citenamefont {Northup},\ and\ \citenamefont
  {Kimble}}]{birnbaum_photon_2005}%
  \BibitemOpen
  \bibfield  {author} {\bibinfo {author} {\bibfnamefont {K.~M.}\ \bibnamefont
  {Birnbaum}}, \bibinfo {author} {\bibfnamefont {A.}~\bibnamefont {Boca}},
  \bibinfo {author} {\bibfnamefont {R.}~\bibnamefont {Miller}}, \bibinfo
  {author} {\bibfnamefont {A.~D.}\ \bibnamefont {Boozer}}, \bibinfo {author}
  {\bibfnamefont {T.~E.}\ \bibnamefont {Northup}},\ and\ \bibinfo {author}
  {\bibfnamefont {H.~J.}\ \bibnamefont {Kimble}},\ }\bibfield  {title}
  {\bibinfo {title} {Photon blockade in an optical cavity with one trapped
  atom},\ }\href {https://doi.org/10.1038/nature03804} {\bibfield  {journal}
  {\bibinfo  {journal} {Nature}\ }\textbf {\bibinfo {volume} {436}},\ \bibinfo
  {pages} {87} (\bibinfo {year} {2005})}\BibitemShut {NoStop}%
\bibitem [{\citenamefont {Yang}\ \emph {et~al.}(2017)\citenamefont {Yang},
  \citenamefont {Pelton}, \citenamefont {Bodnarchuk}, \citenamefont
  {Kovalenko},\ and\ \citenamefont {Waks}}]{yang_spontaneous_2017}%
  \BibitemOpen
  \bibfield  {author} {\bibinfo {author} {\bibfnamefont {Z.}~\bibnamefont
  {Yang}}, \bibinfo {author} {\bibfnamefont {M.}~\bibnamefont {Pelton}},
  \bibinfo {author} {\bibfnamefont {M.~I.}\ \bibnamefont {Bodnarchuk}},
  \bibinfo {author} {\bibfnamefont {M.~V.}\ \bibnamefont {Kovalenko}},\ and\
  \bibinfo {author} {\bibfnamefont {E.}~\bibnamefont {Waks}},\ }\bibfield
  {title} {\bibinfo {title} {Spontaneous emission enhancement of colloidal
  perovskite nanocrystals by a photonic crystal cavity},\ }\href
  {https://doi.org/10.1063/1.5000248} {\bibfield  {journal} {\bibinfo
  {journal} {Applied Physics Letters}\ }\textbf {\bibinfo {volume} {111}},\
  \bibinfo {pages} {221104} (\bibinfo {year} {2017})}\BibitemShut {NoStop}%
\bibitem [{\citenamefont {Jun}\ \emph {et~al.}(2024)\citenamefont {Jun},
  \citenamefont {Kim}, \citenamefont {Choi}, \citenamefont {Kim}, \citenamefont
  {Park}, \citenamefont {Kim}, \citenamefont {Shin},\ and\ \citenamefont
  {Cho}}]{jun_ultrafast_2023}%
  \BibitemOpen
  \bibfield  {author} {\bibinfo {author} {\bibfnamefont {S.}~\bibnamefont
  {Jun}}, \bibinfo {author} {\bibfnamefont {J.}~\bibnamefont {Kim}}, \bibinfo
  {author} {\bibfnamefont {M.}~\bibnamefont {Choi}}, \bibinfo {author}
  {\bibfnamefont {B.~S.}\ \bibnamefont {Kim}}, \bibinfo {author} {\bibfnamefont
  {J.}~\bibnamefont {Park}}, \bibinfo {author} {\bibfnamefont {D.}~\bibnamefont
  {Kim}}, \bibinfo {author} {\bibfnamefont {B.}~\bibnamefont {Shin}},\ and\
  \bibinfo {author} {\bibfnamefont {Y.-H.}\ \bibnamefont {Cho}},\ }\bibfield
  {title} {\bibinfo {title} {Ultrafast and bright quantum emitters from the
  cavity-coupled single perovskite nanocrystals},\ }\href
  {https://doi.org/10.1021/acsnano.3c06760} {\bibfield  {journal} {\bibinfo
  {journal} {ACS Nano}\ }\textbf {\bibinfo {volume} {18}},\ \bibinfo {pages}
  {1396} (\bibinfo {year} {2024})}\BibitemShut {NoStop}%
\bibitem [{\citenamefont {Purkayastha}\ \emph {et~al.}(2024)\citenamefont
  {Purkayastha}, \citenamefont {Gallagher}, \citenamefont {Jiang},
  \citenamefont {Lee}, \citenamefont {Shen}, \citenamefont {Ginger},\ and\
  \citenamefont {Waks}}]{purkayastha_purcell_2024}%
  \BibitemOpen
  \bibfield  {author} {\bibinfo {author} {\bibfnamefont {P.}~\bibnamefont
  {Purkayastha}}, \bibinfo {author} {\bibfnamefont {S.}~\bibnamefont
  {Gallagher}}, \bibinfo {author} {\bibfnamefont {Y.}~\bibnamefont {Jiang}},
  \bibinfo {author} {\bibfnamefont {C.-M.}\ \bibnamefont {Lee}}, \bibinfo
  {author} {\bibfnamefont {G.}~\bibnamefont {Shen}}, \bibinfo {author}
  {\bibfnamefont {D.}~\bibnamefont {Ginger}},\ and\ \bibinfo {author}
  {\bibfnamefont {E.}~\bibnamefont {Waks}},\ }\bibfield  {title} {\bibinfo
  {title} {Purcell enhanced emission and saturable absorption of cavity-coupled
  {CsPbBr}$_{\textrm{3}}$ quantum dots},\ }\href
  {https://doi.org/10.1021/acsphotonics.3c01847} {\bibfield  {journal}
  {\bibinfo  {journal} {ACS Photonics}\ }\textbf {\bibinfo {volume} {11}},\
  \bibinfo {pages} {1638} (\bibinfo {year} {2024})}\BibitemShut {NoStop}%
\bibitem [{\citenamefont {Jeantet}\ \emph {et~al.}(2016)\citenamefont
  {Jeantet}, \citenamefont {Chassagneux}, \citenamefont {Raynaud},
  \citenamefont {Roussignol}, \citenamefont {Lauret}, \citenamefont {Besga},
  \citenamefont {Est{\`e}ve}, \citenamefont {Reichel},\ and\ \citenamefont
  {Voisin}}]{jeantetWidelyTunableSinglePhoton2016b}%
  \BibitemOpen
  \bibfield  {author} {\bibinfo {author} {\bibfnamefont {A.}~\bibnamefont
  {Jeantet}}, \bibinfo {author} {\bibfnamefont {Y.}~\bibnamefont
  {Chassagneux}}, \bibinfo {author} {\bibfnamefont {C.}~\bibnamefont
  {Raynaud}}, \bibinfo {author} {\bibfnamefont {P.}~\bibnamefont {Roussignol}},
  \bibinfo {author} {\bibfnamefont {J.}~\bibnamefont {Lauret}}, \bibinfo
  {author} {\bibfnamefont {B.}~\bibnamefont {Besga}}, \bibinfo {author}
  {\bibfnamefont {J.}~\bibnamefont {Est{\`e}ve}}, \bibinfo {author}
  {\bibfnamefont {J.}~\bibnamefont {Reichel}},\ and\ \bibinfo {author}
  {\bibfnamefont {C.}~\bibnamefont {Voisin}},\ }\bibfield  {title} {\bibinfo
  {title} {Widely {Tunable} {Single}-{Photon} {Source} from a {Carbon}
  {Nanotube} in the {Purcell} {Regime}},\ }\href
  {https://doi.org/10.1103/PhysRevLett.116.247402} {\bibfield  {journal}
  {\bibinfo  {journal} {Physical Review Letters}\ }\textbf {\bibinfo {volume}
  {116}},\ \bibinfo {pages} {247402} (\bibinfo {year} {2016})}\BibitemShut
  {NoStop}%
\bibitem [{\citenamefont {Borel}\ \emph {et~al.}(2023)\citenamefont {Borel},
  \citenamefont {Habrant-Claude}, \citenamefont {Rapisarda}, \citenamefont
  {Reichel}, \citenamefont {Doorn}, \citenamefont {Voisin},\ and\ \citenamefont
  {Chassagneux}}]{borel_telecom_2023}%
  \BibitemOpen
  \bibfield  {author} {\bibinfo {author} {\bibfnamefont {A.}~\bibnamefont
  {Borel}}, \bibinfo {author} {\bibfnamefont {T.}~\bibnamefont
  {Habrant-Claude}}, \bibinfo {author} {\bibfnamefont {F.}~\bibnamefont
  {Rapisarda}}, \bibinfo {author} {\bibfnamefont {J.}~\bibnamefont {Reichel}},
  \bibinfo {author} {\bibfnamefont {S.~K.}\ \bibnamefont {Doorn}}, \bibinfo
  {author} {\bibfnamefont {C.}~\bibnamefont {Voisin}},\ and\ \bibinfo {author}
  {\bibfnamefont {Y.}~\bibnamefont {Chassagneux}},\ }\bibfield  {title}
  {\bibinfo {title} {Telecom {Band} {Single}-{Photon} {Source} {Using} a
  {Grafted} {Carbon} {Nanotube} {Coupled} to a {Fiber} {Fabry}-{Perot} {Cavity}
  in the {Purcell} {Regime}},\ }\href
  {https://doi.org/10.1021/acsphotonics.3c00541} {\bibfield  {journal}
  {\bibinfo  {journal} {ACS Photonics}\ }\textbf {\bibinfo {volume} {10}},\
  \bibinfo {pages} {2839} (\bibinfo {year} {2023})}\BibitemShut {NoStop}%
\bibitem [{\citenamefont {Riedel}\ \emph {et~al.}(2017)\citenamefont {Riedel},
  \citenamefont {S{\"o}llner}, \citenamefont {Shields}, \citenamefont
  {Starosielec}, \citenamefont {Appel}, \citenamefont {Neu}, \citenamefont
  {Maletinsky},\ and\ \citenamefont {Warburton}}]{riedel_deterministic_2017}%
  \BibitemOpen
  \bibfield  {author} {\bibinfo {author} {\bibfnamefont {D.}~\bibnamefont
  {Riedel}}, \bibinfo {author} {\bibfnamefont {I.}~\bibnamefont {S{\"o}llner}},
  \bibinfo {author} {\bibfnamefont {B.~J.}\ \bibnamefont {Shields}}, \bibinfo
  {author} {\bibfnamefont {S.}~\bibnamefont {Starosielec}}, \bibinfo {author}
  {\bibfnamefont {P.}~\bibnamefont {Appel}}, \bibinfo {author} {\bibfnamefont
  {E.}~\bibnamefont {Neu}}, \bibinfo {author} {\bibfnamefont {P.}~\bibnamefont
  {Maletinsky}},\ and\ \bibinfo {author} {\bibfnamefont {R.~J.}\ \bibnamefont
  {Warburton}},\ }\bibfield  {title} {\bibinfo {title} {Deterministic
  {Enhancement} of {Coherent} {Photon} {Generation} from a {Nitrogen}-{Vacancy}
  {Center} in {Ultrapure} {Diamond}},\ }\href
  {https://doi.org/10.1103/PhysRevX.7.031040} {\bibfield  {journal} {\bibinfo
  {journal} {Physical Review X}\ }\textbf {\bibinfo {volume} {7}},\ \bibinfo
  {pages} {031040} (\bibinfo {year} {2017})}\BibitemShut {NoStop}%
\bibitem [{\citenamefont {Wang}\ \emph {et~al.}(2019)\citenamefont {Wang},
  \citenamefont {Kelkar}, \citenamefont {Martin-Cano}, \citenamefont
  {Rattenbacher}, \citenamefont {Shkarin}, \citenamefont {Utikal},
  \citenamefont {Götzinger},\ and\ \citenamefont
  {Sandoghdar}}]{wang_turning_2019}%
  \BibitemOpen
  \bibfield  {author} {\bibinfo {author} {\bibfnamefont {D.}~\bibnamefont
  {Wang}}, \bibinfo {author} {\bibfnamefont {H.}~\bibnamefont {Kelkar}},
  \bibinfo {author} {\bibfnamefont {D.}~\bibnamefont {Martin-Cano}}, \bibinfo
  {author} {\bibfnamefont {D.}~\bibnamefont {Rattenbacher}}, \bibinfo {author}
  {\bibfnamefont {A.}~\bibnamefont {Shkarin}}, \bibinfo {author} {\bibfnamefont
  {T.}~\bibnamefont {Utikal}}, \bibinfo {author} {\bibfnamefont
  {S.}~\bibnamefont {Götzinger}},\ and\ \bibinfo {author} {\bibfnamefont
  {V.}~\bibnamefont {Sandoghdar}},\ }\bibfield  {title} {\bibinfo {title}
  {Turning a molecule into a coherent two-level quantum system},\ }\href
  {https://doi.org/10.1038/s41567-019-0436-5} {\bibfield  {journal} {\bibinfo
  {journal} {Nature Physics}\ }\textbf {\bibinfo {volume} {15}},\ \bibinfo
  {pages} {483} (\bibinfo {year} {2019})}\BibitemShut {NoStop}%
\bibitem [{\citenamefont {Najer}\ \emph {et~al.}(2019)\citenamefont {Najer},
  \citenamefont {S{\"o}llner}, \citenamefont {Sekatski}, \citenamefont
  {Dolique}, \citenamefont {L{\"o}bl}, \citenamefont {Riedel}, \citenamefont
  {Schott}, \citenamefont {Starosielec}, \citenamefont {Valentin},
  \citenamefont {Wieck}, \citenamefont {Sangouard}, \citenamefont {Ludwig},\
  and\ \citenamefont {Warburton}}]{najer_gated_2019}%
  \BibitemOpen
  \bibfield  {author} {\bibinfo {author} {\bibfnamefont {D.}~\bibnamefont
  {Najer}}, \bibinfo {author} {\bibfnamefont {I.}~\bibnamefont {S{\"o}llner}},
  \bibinfo {author} {\bibfnamefont {P.}~\bibnamefont {Sekatski}}, \bibinfo
  {author} {\bibfnamefont {V.}~\bibnamefont {Dolique}}, \bibinfo {author}
  {\bibfnamefont {M.~C.}\ \bibnamefont {L{\"o}bl}}, \bibinfo {author}
  {\bibfnamefont {D.}~\bibnamefont {Riedel}}, \bibinfo {author} {\bibfnamefont
  {R.}~\bibnamefont {Schott}}, \bibinfo {author} {\bibfnamefont
  {S.}~\bibnamefont {Starosielec}}, \bibinfo {author} {\bibfnamefont {S.~R.}\
  \bibnamefont {Valentin}}, \bibinfo {author} {\bibfnamefont {A.~D.}\
  \bibnamefont {Wieck}}, \bibinfo {author} {\bibfnamefont {N.}~\bibnamefont
  {Sangouard}}, \bibinfo {author} {\bibfnamefont {A.}~\bibnamefont {Ludwig}},\
  and\ \bibinfo {author} {\bibfnamefont {R.~J.}\ \bibnamefont {Warburton}},\
  }\bibfield  {title} {\bibinfo {title} {A gated quantum dot strongly coupled
  to an optical microcavity},\ }\href
  {https://doi.org/10.1038/s41586-019-1709-y} {\bibfield  {journal} {\bibinfo
  {journal} {Nature}\ }\textbf {\bibinfo {volume} {575}},\ \bibinfo {pages}
  {622} (\bibinfo {year} {2019})}\BibitemShut {NoStop}%
\bibitem [{\citenamefont {Farrow}\ \emph {et~al.}(2023)\citenamefont {Farrow},
  \citenamefont {Dhawan}, \citenamefont {Marshall}, \citenamefont {Ghorbal},
  \citenamefont {Son}, \citenamefont {Snaith}, \citenamefont {Smith},\ and\
  \citenamefont {Taylor}}]{farrow_ultranarrow_2023}%
  \BibitemOpen
  \bibfield  {author} {\bibinfo {author} {\bibfnamefont {T.}~\bibnamefont
  {Farrow}}, \bibinfo {author} {\bibfnamefont {A.~R.}\ \bibnamefont {Dhawan}},
  \bibinfo {author} {\bibfnamefont {A.~R.}\ \bibnamefont {Marshall}}, \bibinfo
  {author} {\bibfnamefont {A.}~\bibnamefont {Ghorbal}}, \bibinfo {author}
  {\bibfnamefont {W.}~\bibnamefont {Son}}, \bibinfo {author} {\bibfnamefont
  {H.~J.}\ \bibnamefont {Snaith}}, \bibinfo {author} {\bibfnamefont {J.~M.}\
  \bibnamefont {Smith}},\ and\ \bibinfo {author} {\bibfnamefont {R.~A.}\
  \bibnamefont {Taylor}},\ }\bibfield  {title} {\bibinfo {title} {Ultranarrow
  {Line} {Width} {Room}-{Temperature} {Single}-{Photon} {Source} from
  {Perovskite} {Quantum} {Dot} {Embedded} in {Optical} {Microcavity}},\ }\href
  {https://doi.org/10.1021/acs.nanolett.3c02058} {\bibfield  {journal}
  {\bibinfo  {journal} {Nano Letters}\ }\textbf {\bibinfo {volume} {23}},\
  \bibinfo {pages} {10667} (\bibinfo {year} {2023})}\BibitemShut {NoStop}%
\bibitem [{\citenamefont {Rain{\`o}}\ \emph {et~al.}(2019)\citenamefont
  {Rain{\`o}}, \citenamefont {Landuyt}, \citenamefont {Krieg}, \citenamefont
  {Bernasconi}, \citenamefont {Ochsenbein}, \citenamefont {Dirin},
  \citenamefont {Bodnarchuk},\ and\ \citenamefont
  {Kovalenko}}]{raino_underestimated_2019}%
  \BibitemOpen
  \bibfield  {author} {\bibinfo {author} {\bibfnamefont {G.}~\bibnamefont
  {Rain{\`o}}}, \bibinfo {author} {\bibfnamefont {A.}~\bibnamefont {Landuyt}},
  \bibinfo {author} {\bibfnamefont {F.}~\bibnamefont {Krieg}}, \bibinfo
  {author} {\bibfnamefont {C.}~\bibnamefont {Bernasconi}}, \bibinfo {author}
  {\bibfnamefont {S.~T.}\ \bibnamefont {Ochsenbein}}, \bibinfo {author}
  {\bibfnamefont {D.~N.}\ \bibnamefont {Dirin}}, \bibinfo {author}
  {\bibfnamefont {M.~I.}\ \bibnamefont {Bodnarchuk}},\ and\ \bibinfo {author}
  {\bibfnamefont {M.~V.}\ \bibnamefont {Kovalenko}},\ }\bibfield  {title}
  {\bibinfo {title} {Underestimated {Effect} of a {Polymer} {Matrix} on the
  {Light} {Emission} of {Single} {CsPbBr}$_{\textrm{3}}$ {Nanocrystals}},\
  }\href {https://doi.org/10.1021/acs.nanolett.9b00689} {\bibfield  {journal}
  {\bibinfo  {journal} {Nano Letters}\ }\textbf {\bibinfo {volume} {19}},\
  \bibinfo {pages} {3648} (\bibinfo {year} {2019})}\BibitemShut {NoStop}%
\bibitem [{\citenamefont {Amara}\ \emph {et~al.}(2023)\citenamefont {Amara},
  \citenamefont {Said}, \citenamefont {Huo}, \citenamefont {Pierret},
  \citenamefont {Voisin}, \citenamefont {Gao}, \citenamefont {Xiong},\ and\
  \citenamefont {Diederichs}}]{amara_spectral_2023}%
  \BibitemOpen
  \bibfield  {author} {\bibinfo {author} {\bibfnamefont {M.-R.}\ \bibnamefont
  {Amara}}, \bibinfo {author} {\bibfnamefont {Z.}~\bibnamefont {Said}},
  \bibinfo {author} {\bibfnamefont {C.}~\bibnamefont {Huo}}, \bibinfo {author}
  {\bibfnamefont {A.}~\bibnamefont {Pierret}}, \bibinfo {author} {\bibfnamefont
  {C.}~\bibnamefont {Voisin}}, \bibinfo {author} {\bibfnamefont
  {W.}~\bibnamefont {Gao}}, \bibinfo {author} {\bibfnamefont {Q.}~\bibnamefont
  {Xiong}},\ and\ \bibinfo {author} {\bibfnamefont {C.}~\bibnamefont
  {Diederichs}},\ }\bibfield  {title} {\bibinfo {title} {Spectral {Fingerprint}
  of {Quantum} {Confinement} in {Single} {CsPbBr}$_{\textrm{3}}$
  {Nanocrystals}},\ }\href {https://doi.org/10.1021/acs.nanolett.3c00793}
  {\bibfield  {journal} {\bibinfo  {journal} {Nano Letters}\ }\textbf {\bibinfo
  {volume} {23}},\ \bibinfo {pages} {3607} (\bibinfo {year} {2023})},\ \bibinfo
  {note} {publisher: American Chemical Society}\BibitemShut {NoStop}%
\bibitem [{\citenamefont {Auff\`eves}\ \emph {et~al.}(2010)\citenamefont
  {Auff\`eves}, \citenamefont {Gerace}, \citenamefont {G\'erard}, \citenamefont
  {Santos}, \citenamefont {Andreani},\ and\ \citenamefont
  {Poizat}}]{Auffeves2010}%
  \BibitemOpen
  \bibfield  {author} {\bibinfo {author} {\bibfnamefont {A.}~\bibnamefont
  {Auff\`eves}}, \bibinfo {author} {\bibfnamefont {D.}~\bibnamefont {Gerace}},
  \bibinfo {author} {\bibfnamefont {J.-M.}\ \bibnamefont {G\'erard}}, \bibinfo
  {author} {\bibfnamefont {M.~F. m.~c.}\ \bibnamefont {Santos}}, \bibinfo
  {author} {\bibfnamefont {L.~C.}\ \bibnamefont {Andreani}},\ and\ \bibinfo
  {author} {\bibfnamefont {J.-P.}\ \bibnamefont {Poizat}},\ }\bibfield  {title}
  {\bibinfo {title} {Controlling the dynamics of a coupled atom-cavity system
  by pure dephasing},\ }\href {https://doi.org/10.1103/PhysRevB.81.245419}
  {\bibfield  {journal} {\bibinfo  {journal} {Phys. Rev. B}\ }\textbf {\bibinfo
  {volume} {81}},\ \bibinfo {pages} {245419} (\bibinfo {year}
  {2010})}\BibitemShut {NoStop}%
\bibitem [{Note1()}]{Note1}%
  \BibitemOpen
  \bibinfo {note} {We note that the simple extraction of the dip for $\lambda
  ^3/V\simeq 0.038$ is far from the general trend. However, this discrepancy is
  less critical when considering the full theoretical analysis as shown in
  Figure~\ref {fig:rabi_couplingNEW}b.}\BibitemShut {Stop}%
\bibitem [{\citenamefont {Grange}\ \emph {et~al.}(2015)\citenamefont {Grange},
  \citenamefont {Hornecker}, \citenamefont {Hunger}, \citenamefont {Poizat},
  \citenamefont {G{\'e}rard}, \citenamefont {Senellart},\ and\ \citenamefont
  {Auff{\`e}ves}}]{grange_cavity-funneled_2015}%
  \BibitemOpen
  \bibfield  {author} {\bibinfo {author} {\bibfnamefont {T.}~\bibnamefont
  {Grange}}, \bibinfo {author} {\bibfnamefont {G.}~\bibnamefont {Hornecker}},
  \bibinfo {author} {\bibfnamefont {D.}~\bibnamefont {Hunger}}, \bibinfo
  {author} {\bibfnamefont {J.-P.}\ \bibnamefont {Poizat}}, \bibinfo {author}
  {\bibfnamefont {J.-M.}\ \bibnamefont {G{\'e}rard}}, \bibinfo {author}
  {\bibfnamefont {P.}~\bibnamefont {Senellart}},\ and\ \bibinfo {author}
  {\bibfnamefont {A.}~\bibnamefont {Auff{\`e}ves}},\ }\bibfield  {title}
  {\bibinfo {title} {Cavity-{Funneled} {Generation} of {Indistinguishable}
  {Single} {Photons} from {Strongly} {Dissipative} {Quantum} {Emitters}},\
  }\href {https://doi.org/10.1103/PhysRevLett.114.193601} {\bibfield  {journal}
  {\bibinfo  {journal} {Physical Review Letters}\ }\textbf {\bibinfo {volume}
  {114}},\ \bibinfo {pages} {193601} (\bibinfo {year} {2015})}\BibitemShut
  {NoStop}%
\bibitem [{\citenamefont {Hunger}\ \emph {et~al.}(2010)\citenamefont {Hunger},
  \citenamefont {Steinmetz}, \citenamefont {Colombe}, \citenamefont {Deutsch},
  \citenamefont {H{\"a}nsch},\ and\ \citenamefont
  {Reichel}}]{hunger_fiber_2010}%
  \BibitemOpen
  \bibfield  {author} {\bibinfo {author} {\bibfnamefont {D.}~\bibnamefont
  {Hunger}}, \bibinfo {author} {\bibfnamefont {T.}~\bibnamefont {Steinmetz}},
  \bibinfo {author} {\bibfnamefont {Y.}~\bibnamefont {Colombe}}, \bibinfo
  {author} {\bibfnamefont {C.}~\bibnamefont {Deutsch}}, \bibinfo {author}
  {\bibfnamefont {T.~W.}\ \bibnamefont {H{\"a}nsch}},\ and\ \bibinfo {author}
  {\bibfnamefont {J.}~\bibnamefont {Reichel}},\ }\bibfield  {title} {\bibinfo
  {title} {A fiber {Fabry}-{Perot} cavity with high finesse},\ }\href
  {https://doi.org/10.1088/1367-2630/12/6/065038} {\bibfield  {journal}
  {\bibinfo  {journal} {New Journal of Physics}\ }\textbf {\bibinfo {volume}
  {12}},\ \bibinfo {pages} {065038} (\bibinfo {year} {2010})}\BibitemShut
  {NoStop}%
\end{thebibliography}%


\begin{thebibliography}{7}%
\makeatletter
\providecommand \@ifxundefined [1]{%
 \@ifx{#1\undefined}
}%
\providecommand \@ifnum [1]{%
 \ifnum #1\expandafter \@firstoftwo
 \else \expandafter \@secondoftwo
 \fi
}%
\providecommand \@ifx [1]{%
 \ifx #1\expandafter \@firstoftwo
 \else \expandafter \@secondoftwo
 \fi
}%
\providecommand \natexlab [1]{#1}%
\providecommand \enquote  [1]{``#1''}%
\providecommand \bibnamefont  [1]{#1}%
\providecommand \bibfnamefont [1]{#1}%
\providecommand \citenamefont [1]{#1}%
\providecommand \href@noop [0]{\@secondoftwo}%
\providecommand \href [0]{\begingroup \@sanitize@url \@href}%
\providecommand \@href[1]{\@@startlink{#1}\@@href}%
\providecommand \@@href[1]{\endgroup#1\@@endlink}%
\providecommand \@sanitize@url [0]{\catcode `\\12\catcode `\$12\catcode
  `\&12\catcode `\#12\catcode `\^12\catcode `\_12\catcode `\%12\relax}%
\providecommand \@@startlink[1]{}%
\providecommand \@@endlink[0]{}%
\providecommand \url  [0]{\begingroup\@sanitize@url \@url }%
\providecommand \@url [1]{\endgroup\@href {#1}{\urlprefix }}%
\providecommand \urlprefix  [0]{URL }%
\providecommand \Eprint [0]{\href }%
\providecommand \doibase [0]{https://doi.org/}%
\providecommand \selectlanguage [0]{\@gobble}%
\providecommand \bibinfo  [0]{\@secondoftwo}%
\providecommand \bibfield  [0]{\@secondoftwo}%
\providecommand \translation [1]{[#1]}%
\providecommand \BibitemOpen [0]{}%
\providecommand \bibitemStop [0]{}%
\providecommand \bibitemNoStop [0]{.\EOS\space}%
\providecommand \EOS [0]{\spacefactor3000\relax}%
\providecommand \BibitemShut  [1]{\csname bibitem#1\endcsname}%
\let\auto@bib@innerbib\@empty
%</preamble>
\bibitem [{\citenamefont {Grundmann}\ and\ \citenamefont
  {Bimberg}(1997)}]{Grund_Thoery_1997}%
  \BibitemOpen
  \bibfield  {author} {\bibinfo {author} {\bibfnamefont {M.}~\bibnamefont
  {Grundmann}}\ and\ \bibinfo {author} {\bibfnamefont {D.}~\bibnamefont
  {Bimberg}},\ }\href {https://doi.org/10.1103/PhysRevB.55.9740} {\bibfield
  {journal} {\bibinfo  {journal} {Phys. Rev. B}\ }\textbf {\bibinfo {volume}
  {55}},\ \bibinfo {pages} {9740} (\bibinfo {year} {1997})}\BibitemShut
  {NoStop}%
\bibitem [{\citenamefont {Protesescu}\ \emph {et~al.}(2015)\citenamefont
  {Protesescu}, \citenamefont {Yakunin}, \citenamefont {Bodnarchuk},
  \citenamefont {Krieg}, \citenamefont {Caputo}, \citenamefont {Hendon},
  \citenamefont {Yang}, \citenamefont {Walsh},\ and\ \citenamefont
  {Kovalenko}}]{Protesescu}%
  \BibitemOpen
  \bibfield  {author} {\bibinfo {author} {\bibfnamefont {L.}~\bibnamefont
  {Protesescu}}, \bibinfo {author} {\bibfnamefont {S.}~\bibnamefont {Yakunin}},
  \bibinfo {author} {\bibfnamefont {M.~I.}\ \bibnamefont {Bodnarchuk}},
  \bibinfo {author} {\bibfnamefont {F.}~\bibnamefont {Krieg}}, \bibinfo
  {author} {\bibfnamefont {R.}~\bibnamefont {Caputo}}, \bibinfo {author}
  {\bibfnamefont {C.~H.}\ \bibnamefont {Hendon}}, \bibinfo {author}
  {\bibfnamefont {R.~X.}\ \bibnamefont {Yang}}, \bibinfo {author}
  {\bibfnamefont {A.}~\bibnamefont {Walsh}},\ and\ \bibinfo {author}
  {\bibfnamefont {M.~V.}\ \bibnamefont {Kovalenko}},\ }\href
  {https://doi.org/10.1021/nl5048779} {\bibfield  {journal} {\bibinfo
  {journal} {Nano Letters}\ }\textbf {\bibinfo {volume} {15}},\ \bibinfo
  {pages} {3692} (\bibinfo {year} {2015})}\BibitemShut {NoStop}%
\bibitem [{\citenamefont {Bo}\ \emph {et~al.}(2021)\citenamefont {Bo},
  \citenamefont {Sun}, \citenamefont {Wan}, \citenamefont {Huang},
  \citenamefont {Chen}, \citenamefont {Chen}, \citenamefont {Li}, \citenamefont
  {Shen}, \citenamefont {Li}, \citenamefont {Xia}, \citenamefont {Ye},
  \citenamefont {Chen},\ and\ \citenamefont {Chen}}]{bo_perovskite_2021}%
  \BibitemOpen
  \bibfield  {author} {\bibinfo {author} {\bibfnamefont {J.}~\bibnamefont
  {Bo}}, \bibinfo {author} {\bibfnamefont {X.}~\bibnamefont {Sun}}, \bibinfo
  {author} {\bibfnamefont {P.}~\bibnamefont {Wan}}, \bibinfo {author}
  {\bibfnamefont {D.}~\bibnamefont {Huang}}, \bibinfo {author} {\bibfnamefont
  {X.}~\bibnamefont {Chen}}, \bibinfo {author} {\bibfnamefont {M.}~\bibnamefont
  {Chen}}, \bibinfo {author} {\bibfnamefont {R.}~\bibnamefont {Li}}, \bibinfo
  {author} {\bibfnamefont {D.}~\bibnamefont {Shen}}, \bibinfo {author}
  {\bibfnamefont {Q.}~\bibnamefont {Li}}, \bibinfo {author} {\bibfnamefont
  {W.}~\bibnamefont {Xia}}, \bibinfo {author} {\bibfnamefont {Z.}~\bibnamefont
  {Ye}}, \bibinfo {author} {\bibfnamefont {Y.}~\bibnamefont {Chen}},\ and\
  \bibinfo {author} {\bibfnamefont {S.}~\bibnamefont {Chen}},\ }\href
  {https://doi.org/10.1021/acs.jpclett.1c02472} {\bibfield  {journal} {\bibinfo
   {journal} {The Journal of Physical Chemistry Letters}\ }\textbf {\bibinfo
  {volume} {12}},\ \bibinfo {pages} {9115} (\bibinfo {year}
  {2021})}\BibitemShut {NoStop}%
\bibitem [{\citenamefont {Reed}\ \emph {et~al.}(1987)\citenamefont {Reed},
  \citenamefont {Giergiel}, \citenamefont {Hemminger},\ and\ \citenamefont
  {Ushioda}}]{reed_dipole_1987}%
  \BibitemOpen
  \bibfield  {author} {\bibinfo {author} {\bibfnamefont {C.~E.}\ \bibnamefont
  {Reed}}, \bibinfo {author} {\bibfnamefont {J.}~\bibnamefont {Giergiel}},
  \bibinfo {author} {\bibfnamefont {J.~C.}\ \bibnamefont {Hemminger}},\ and\
  \bibinfo {author} {\bibfnamefont {S.}~\bibnamefont {Ushioda}},\ }\href
  {https://doi.org/10.1103/PhysRevB.36.4990} {\bibfield  {journal} {\bibinfo
  {journal} {Phys. Rev. B}\ }\textbf {\bibinfo {volume} {36}},\ \bibinfo
  {pages} {4990} (\bibinfo {year} {1987})}\BibitemShut {NoStop}%
\bibitem [{\citenamefont {Amara}\ \emph {et~al.}(2024)\citenamefont {Amara},
  \citenamefont {Huo}, \citenamefont {Voisin}, \citenamefont {Xiong},\ and\
  \citenamefont {Diederichs}}]{amara_impact_2024}%
  \BibitemOpen
  \bibfield  {author} {\bibinfo {author} {\bibfnamefont {M.-R.}\ \bibnamefont
  {Amara}}, \bibinfo {author} {\bibfnamefont {C.}~\bibnamefont {Huo}}, \bibinfo
  {author} {\bibfnamefont {C.}~\bibnamefont {Voisin}}, \bibinfo {author}
  {\bibfnamefont {Q.}~\bibnamefont {Xiong}},\ and\ \bibinfo {author}
  {\bibfnamefont {C.}~\bibnamefont {Diederichs}},\ }\href
  {https://doi.org/10.1021/acs.nanolett.4c00605} {\bibfield  {journal}
  {\bibinfo  {journal} {Nano Letters}\ }\textbf {\bibinfo {volume} {24}},\
  \bibinfo {pages} {4265} (\bibinfo {year} {2024})}\BibitemShut {NoStop}%
\bibitem [{\citenamefont {Auff{\`e}ves}\ \emph {et~al.}(2008)\citenamefont
  {Auff{\`e}ves}, \citenamefont {Besga}, \citenamefont {G{\'e}rard},\ and\
  \citenamefont {Poizat}}]{Auffeves2008}%
  \BibitemOpen
  \bibfield  {author} {\bibinfo {author} {\bibfnamefont {A.}~\bibnamefont
  {Auff{\`e}ves}}, \bibinfo {author} {\bibfnamefont {B.}~\bibnamefont {Besga}},
  \bibinfo {author} {\bibfnamefont {J.-M.}\ \bibnamefont {G{\'e}rard}},\ and\
  \bibinfo {author} {\bibfnamefont {J.-P.}\ \bibnamefont {Poizat}},\ }\href
  {https://doi.org/10.1103/PhysRevA.77.063833} {\bibfield  {journal} {\bibinfo
  {journal} {Phys. Rev. A}\ }\textbf {\bibinfo {volume} {77}},\ \bibinfo
  {pages} {063833} (\bibinfo {year} {2008})}\BibitemShut {NoStop}%
\bibitem [{\citenamefont {Auff\`eves}\ \emph {et~al.}(2010)\citenamefont
  {Auff\`eves}, \citenamefont {Gerace}, \citenamefont {G\'erard}, \citenamefont
  {Santos}, \citenamefont {Andreani},\ and\ \citenamefont
  {Poizat}}]{Auffeves2010}%
  \BibitemOpen
  \bibfield  {author} {\bibinfo {author} {\bibfnamefont {A.}~\bibnamefont
  {Auff\`eves}}, \bibinfo {author} {\bibfnamefont {D.}~\bibnamefont {Gerace}},
  \bibinfo {author} {\bibfnamefont {J.-M.}\ \bibnamefont {G\'erard}}, \bibinfo
  {author} {\bibfnamefont {M.~F. m.~c.}\ \bibnamefont {Santos}}, \bibinfo
  {author} {\bibfnamefont {L.~C.}\ \bibnamefont {Andreani}},\ and\ \bibinfo
  {author} {\bibfnamefont {J.-P.}\ \bibnamefont {Poizat}},\ }\href
  {https://doi.org/10.1103/PhysRevB.81.245419} {\bibfield  {journal} {\bibinfo
  {journal} {Phys. Rev. B}\ }\textbf {\bibinfo {volume} {81}},\ \bibinfo
  {pages} {245419} (\bibinfo {year} {2010})}\BibitemShut {NoStop}%
\end{thebibliography}%

\section*{Methods}

\subsection*{Perovskite quantum dot synthesis and sample preparation}
Cesium lead bromide (CsPbBr$_3$) quantum dots were synthesized using a first-generation hot injection method as described by Protesescu et al. \cite{Protesescu}, which consists of two main steps: the preparation of Cs-oleate and its subsequent reaction with a bromide source. Cesium carbonate (Cs$_2$CO$_3$), octadecene (ODE) and oleic acid (OA) were introduced into a reaction flask. The mixture was dried at 120~\si{\degreeCelsius} for 1 hour and then heated to 150~\si{\degreeCelsius} under a nitrogen atmosphere until all the Cs$_2$CO$_3$ had reacted to form the Cs oleate precursor. ODE and lead bromide (PbBr$_2$) were injected into a separate flask and dried under vacuum at 120~\si{\degreeCelsius} for 1 hour. Dried oleylamine (OLA) and OA were then added at 120~\si{\degreeCelsius} under nitrogen. When the PbBr$_2$ was completely dissolved, the temperature was raised to 180~\si{\degreeCelsius}. At this point the Cs-oleate precursor solution was rapidly injected into the reaction mixture. After brief heating, the reaction mixture was rapidly cooled in an ice-water bath. The product was purified by a two-step centrifugation. The resulting CsPbBr$_3$ quantum dots were re-dispersed in toluene to form long-term stable dispersion. The CsPbBr$_3$ quantum dots can be stored for weeks to months if kept in a concentrated solution under inert atmosphere.

\subsection*{Sample preparation for single perovskite quantum dot studies}
For single perovskite quantum dot spectroscopy studies, the synthesized solution was diluted approximately 1:10000 in toluene containing 3~\si{\percent} by mass of polystyrene and spin-coated at 3000~rpm for 70~\si{\second} on a highly reflective planar dielectric mirror. After solvent evaporation, the resulting sample consists of highly dispersed and randomly oriented perovskite quantum dots with typical densities of 0.01~\si{\micro\meter^{-2}} embedded in a 200~\si{\nano\meter}-thick polystyrene film.

\subsection*{Design of the tunable Fabry-Pérot fiber microcavity} 
The plano-concave Fabry-Pérot cavity consists of a large planar dielectric mirror and a concave fiber dielectric mirror, the shape of which has been micro-machined by CO$_2$ laser ablation, resulting in a radius of curvature of about 10~\si{\micro\meter} \cite{hunger_fiber_2010}. Both the planar substrate and the fiber tip were coated by the company Laseroptik with precisely defined dielectric stacks of $\text{SiO}_2 \ / \ \text{Ta}_2\text{O}_5$ to ensure a high reflection coefficient at the emission wavelength of the CsPbBr$_{3}$ quantum dots ($R\simeq 0.9995$ in the 475-585~\si{\nano\meter} range) and a high transmission of the excitation laser ($T\simeq0.9$ in the 430-455~\si{\nano\meter} range). This results in a high cavity finesse $\mathcal{F}\simeq1500$ with a quality factor $Q= 25000-50000$ depending on the longitudinal order $p$ of the cavity mode. The lateral and longitudinal displacements of one mirror with respect to the other needed for spatial and spectral matching are obtained via nanopositioning systems (Attocube, ANPx51, ANPz51).

\subsection*{Low-temperature micro-photoluminescence spectroscopy} 
All optical measurements were performed at cryogenic temperatures using a closed-cycle liquid helium cryostat (Montana Instruments, Cryostation S50). The micro-photoluminescence spectroscopy experiments were performed with a home-built scanning confocal microscope as shown in Figure \ref{fig:setup_scheme}. Excitation was performed through the backside of the planar dielectric mirror on which the perovskite quantum dots are deposited by either a $\sim450$~\si{\nano\meter} continuous wave laser diode (Thorlabs LP450-SF25) or a tunable femtosecond pulsed Ti:Sa laser (Spectra-Physics, Tsunami) frequency doubled with a BBO crystal to operate at 450~\si{\nano\meter}. The photoluminescence was collected by the movable 0.7 numerical aperture aspheric lens in free space configuration, and from the back of the planar mirror or the output port of the machined fiber in cavity configuration. The fiber used for the microcavity is inserted into this lens through a $\sim300$~\si{\micro\meter} wide hole drilled at the lens center to switch between free space and cavity configurations by simple translation of the lens. Outside the cryostat, the photoluminescence was sorted from the excitation laser using a dichroic mirror (Thorlabs, DMLP505) and a long pass filter. It was then dispersed using a grating monochromator (Princeton Instruments HRS-500) and detected by a CCD camera (Teledyne Pixis 100) for spectral envelope studies (with a spectral resolution of 200~\si{\micro\electronvolt}). For time-resolved PL, a 35~\si{\pico\second} resolution single photon avalanche detector (MPD) connected to a correlation acquisition card (Picoquant PicoHarp) was used to record the photon arrival times. The intensity auto-correlation measurements were performed using a Hanbury-Brown and Twiss setup where the photoluminescence from a single perovskite quantum dot was split by a 50/50 beam splitter and the two outputs were directed to two 35~\si{\pico\second} resolution single photon avalanche detectors (MPD) connected to the correlation acquisition card.

\subsection*{Envelope dip value}
The spectral envelope dip value is defined as $dip\equiv\frac{min}{(max_1+max_2)/2}$, where $max_1$ and $max_2$ are the two maxima and $min$ is the central minimum (the three extrema values are obtained by a local second order polynomial fit). To correct for the effect of the different vibration amplitudes between the contact mode $p=14$ and other modes $p\geq 15$, the normalized dip value is multiplied by $\exp(-(\Delta/2)^2 / (2 \sigma_{\text{vib}}^2))$, where $\Delta$ is the doublet splitting and $\sigma_{\text{vib}}$ is the vibration energy variance of the corresponding mode as derived from the white lamp cavity transmission. 

\section*{Acknowledgments}
This work was supported by the French National Research Agency (ANR) through the projects IPER-Nano2 (ANR-18-CE30-0023) and DELICACY (ANR-22-CE47-0001). The authors thank Pascal Morfin, Arnaud Leclercq, and the LPENS mechanical workshop for their assistance in designing and fabricating mechanical components. We also thank Aur\'elie Pierret for her help with the chemical synthesis setup at LPENS and Torben Pöpplau for assistance with fiber micromachining at LKB. Finally, we thank Gabriel H\'etet and Emmanuel Baudin for fruitful discussions.

\section*{Author contributions}
Z.S. and M.R.A. synthesized the pQDs and performed single pQD spectroscopy. Z.S. and A.B. designed and implemented the microcavity setup under the supervision of C.D., C.V., Y.C. and J.R. The cQED experiments were performed by Z.S. and M.C.T. under the supervision of C.D. and Y.C. The model was developed by Y.C. with the input of all co-authors. The paper was written by C.D., Y.C., C.V., Z.S. and M.C.T. with the contribution of all co-authors. C.D. and Y.C. conceived the project and supervised the work. All authors discussed the results and commented on the paper.

\section*{Competing interests}
The authors declare no competing interests.

\section*{Additional information}

%TC:endignore
\end{document}

%\preprint{APS/123-QED}

\title{Supplementary Information : Cavity quantum electrodynamics with single perovskite quantum dots}% Force line breaks with \\
\author{Zakaria Said}
\affiliation{%
 Laboratoire de Physique de l’Ecole Normale Supérieure, ENS, Université PSL, CNRS, Sorbonne Université, Université Paris Cité, F-75005 Paris, France\\
}%
 %\altaffiliation[Also at ]{Laboratoire de Physique de l’ENS, Université PSL, CNRS, Sorbonne Université, Université Paris Cité, 75005 Paris, France}
 \author{Marina Cagnon Trouche}
 \affiliation{%
 Laboratoire de Physique de l’Ecole Normale Supérieure, ENS, Université PSL, CNRS, Sorbonne Université, Université Paris Cité, F-75005 Paris, France\\
}%

 \author{Antoine Borel}
 \affiliation{%
 Laboratoire de Physique de l’Ecole Normale Supérieure, ENS, Université PSL, CNRS, Sorbonne Université, Université Paris Cité, F-75005 Paris, France\\
}%

 %\altaffiliation[]{}
 \author{Mohamed-Raouf Amara}
 \affiliation{%
 Laboratoire de Physique de l’Ecole Normale Supérieure, ENS, Université PSL, CNRS, Sorbonne Université, Université Paris Cité, F-75005 Paris, France\\
}%
 %\altaffiliation[Also at ]{Physics Department, XYZ University.}

 %\altaffiliation[Also at ]{Physics Department, XYZ University.}
  \author{Jakob Reichel}
  \affiliation{%
Laboratoire Kastler Brossel, Sorbonne Université, CNRS, ENS - Université PSL, Collège de France, Paris F-75252, France\\
}
 %\altaffiliation[Also at ]{Physics Department, XYZ University.}
\author{Christophe Voisin}
\affiliation{%
 Laboratoire de Physique de l’Ecole Normale Supérieure, ENS, Université PSL, CNRS, Sorbonne Université, Université Paris Cité, F-75005 Paris, France\\
}%

 %\altaffiliation[Also at ]{Physics Department, XYZ University.}
\author{Carole Diederichs}%
 \email{carole.diederichs@phys.ens.fr}
 
\affiliation{%
 Laboratoire de Physique de l’Ecole Normale Supérieure, ENS, Université PSL, CNRS, Sorbonne Université, Université Paris Cité, F-75005 Paris, France\\
}%
\affiliation{Institut Universitaire de France (IUF), 75231 Paris, France}
 %\altaffiliation[Also at ]{Physics Department, XYZ University.}
 \author{Yannick Chassagneux}%
 \email{yannick.chassagneux@phys.ens.fr}
 \affiliation{%
 Laboratoire de Physique de l’Ecole Normale Supérieure, ENS, Université PSL, CNRS, Sorbonne Université, Université Paris Cité, F-75005 Paris, France\\
}%


%\affiliation{
%\affiliation{
%\affiliation{%


\date{\today}% It is always \today, today,

\maketitle

\section{Microcavity mirrors specifications}
\begin{figure}[H]
    \centering
    \includegraphics{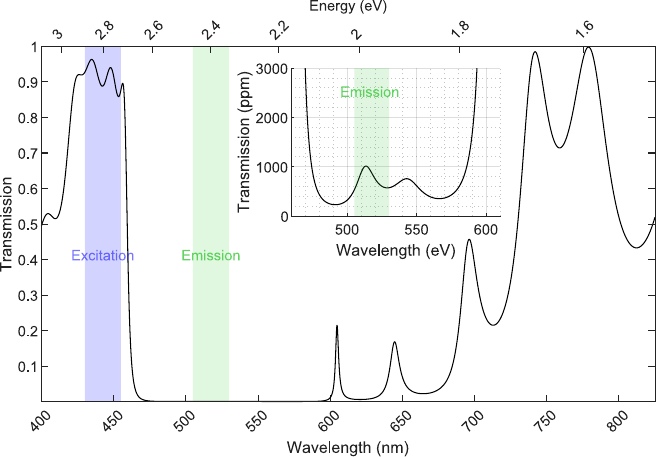}
    \caption{Transmission of the dielectric mirrors of the fibered cavity, where $T>0.85$ in the excitation wavelength range and $T<1000$~ppm in the pQDs emission wavelength range.}
    \label{fig:stopband}
\end{figure}

The dielectric mirrors, the transmission of which is plotted in Figure~\ref{fig:stopband}, have been designed to achieve a high transmission coefficient $T>0.85$ in the excitation wavelength range (blue region), and a flat and high reflectance between $475~\si{nm}$ and $585~\si{nm}$ such that $T<1000$~ppm covers the emission wavelength range of CsPbBr$_3$ lead halide perovskite quantum dots (pQD) (green region). The fibered and planar mirrors used to form the cavity have the same dielectric coating. 

\section{Perovskite quantum dots optical properties}
\subsection{Exciton, trion, and biexciton in the photoluminescence spectrum \label{paragraph:XtrionXX}}
\begin{figure}[!ht]
    %\centering
    \includegraphics{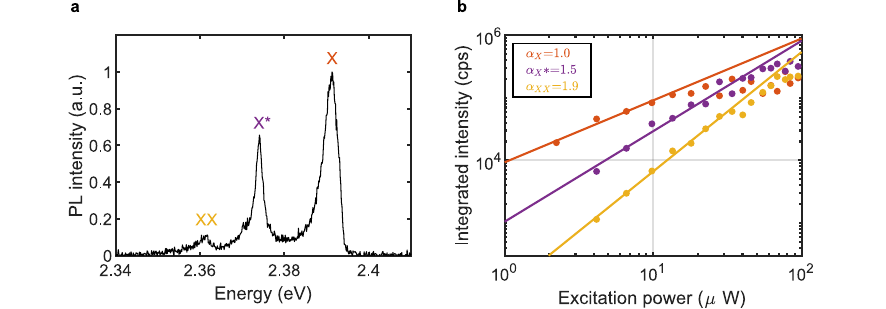}
    \caption{(a) Free space PL spectrum of pQD2 at 10K, showing the exciton (X), trion (X*) and biexciton (XX) emission lines. (b) Excitation power dependence of the intensity of the three spectral lines.}
    \label{fig:full_spectrum_and_slopes}
\end{figure}

The photoluminescence (PL) emission spectrum of pQD2 at cryogenic temperatures (10~\si{\kelvin}) is shown in Figure~\ref{fig:full_spectrum_and_slopes}a, where three main lines can be distinguished. To identify the different excitonic complexes that contribute to the emission spectrum, the power dependence of each line intensity is studied and plotted on a logarithmic scale in Figure~\ref{fig:full_spectrum_and_slopes}b. The PL intensity is fitted to a power law $I\propto P^\alpha$. We obtain $\alpha=$ 1.0, 1.5, 1.9 from the higher to the lower energy peak. Following a random population model \cite{Grund_Thoery_1997}, where $\alpha$ is expected to be 1, 1.5 and 2 for the exciton, trion (with a photo-created charge and not a resident one) and biexciton respectively, we can assign the three main lines accordingly.

\subsection{Alternating exciton and trion emission \label{paragraph:altXtrion}}
\begin{figure}[!ht]
    %\centering
    \includegraphics{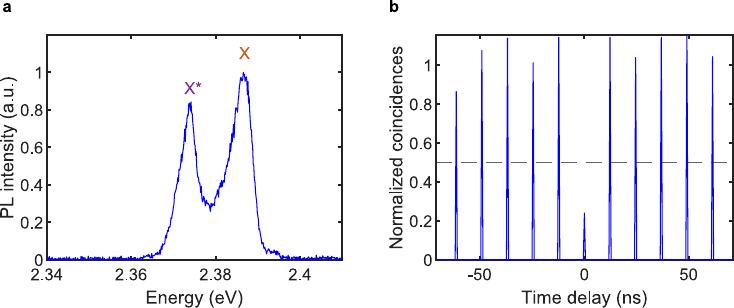}
    \caption{(a) Filtered spectrum of pQD2 where only the exciton ($X$) and the trion ($X^*$) emission lines are detected. (b) Second-order intensity correlation function measured on the integrated spectrum (a) where $g^{(2)}(0)<0.5$.}
    \label{fig:X_and_Xst_spectrum_and_g2}
\end{figure}

We verify that the emission from the exciton ($X$) and the trion ($X^*$) cannot occur simultaneously, as they originate from either a neutral or a charged pQD. To do so, the emission statistic is analyzed by measuring the second-order intensity correlation function ($g^{(2)}$) with a Hanbury-Brown and Twiss experiment under pulsed excitation. This measurement is performed on the filtered spectrum shown in Figure~\ref{fig:X_and_Xst_spectrum_and_g2}a, where the excitation power has been chosen so that both the $X$ and $X^*$ lines exhibit comparable intensities, while the other spectral contributions (biexciton) have been filtered out. As shown in Figure~\ref{fig:X_and_Xst_spectrum_and_g2}b, a photon antibunching with $g^{(2)}(0)<0.5$ is observed, confirming that the pQD is a single photon emitter and that the emissions from the exciton and the trion are not simultaneous.

\begin{figure}[!ht]
    %\centering    
    \includegraphics{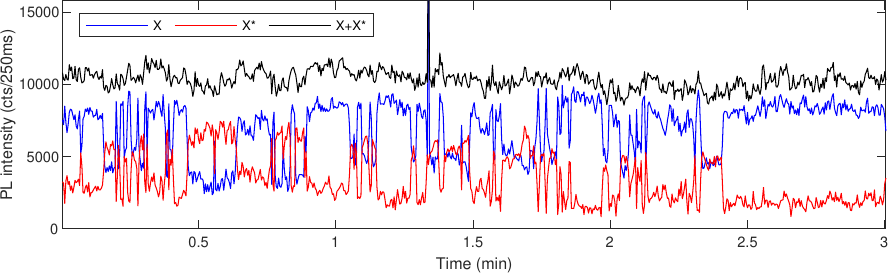}
    \caption{Intensity temporal traces over 3~\si{\minute} of the exciton ($X$, blue), the trion ($X^*$, red) and their sum ($X+X^*$, black) for pQD3.}
    \label{fig:Timetraces}
\end{figure}

The signature of a non-simultaneous emission from $X$ and $X^*$ can also be observed on a longer time scale in their intensity temporal traces, as displayed in Figure~\ref{fig:Timetraces} for pQD3. These time traces show anticorrelations between $X$ and $X^*$, but a stable intensity is retrieved when considering the sum of both temporal traces, with a resulting intensity corresponding to the overall average intensity of the pQD emission. We conclude that the emission from the exciton and the trion are mutually exclusive, and that their transitions are characterized by similar radiative quantum yields.   

\subsection{Stability of the excitonic emission}
\begin{figure}[!ht]
    %\centering
    \includegraphics{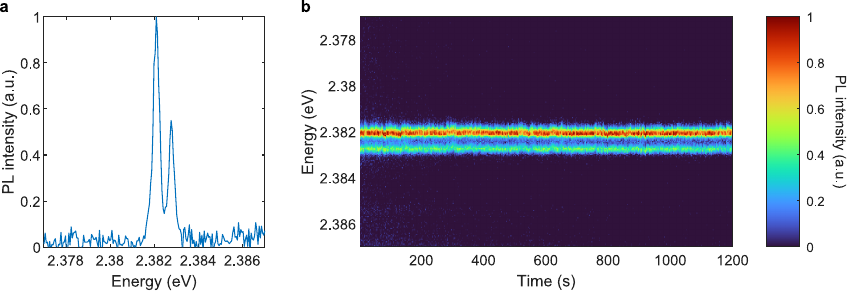}
    \caption{(a) Free space PL spectrum of pQD1 showing the emission of an excitonic doublet with an acquisition time of $2.5$~\si{s}. (b) Temporal trace over $20$~\si{\minute} of the pQD excitonic doublet. The acquisition time for each spectrum is $2.5$~\si{s}.}
    \label{fig:stableXdoublet}
\end{figure}

\begin{figure}[!ht]
    %\centering
    \includegraphics{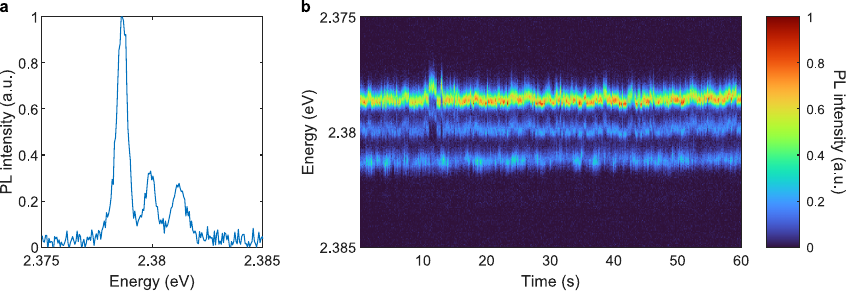}
    \caption{(a) Free space PL spectrum of a pQD showing the emission of an excitonic triplet with an acquisition time of $100$~\si{\milli s}. (b) Temporal trace over $60$~\si{s} of the pQD excitonic triplet. The acquisition time for each spectrum is $100$~\si{\milli s}.}
    \label{fig:stableXtriplet}
\end{figure}

We also highlight that the emission of the neutral exciton is stable over time. Figure~\ref{fig:stableXdoublet}b presents the time evolution over $20$~\si{\min} of the exciton doublet of pQD1 studied in the main text. The spectrum is reproduced in Figure~\ref{fig:stableXdoublet}a. Only weak and correlated spectral and intensity fluctuations can be observed. Another example of a temporal trace over $60$~\si{\second} is shown in Figure~\ref{fig:stableXtriplet}b for a pQD exhibiting an excitonic triplet in its spectrum (Figure~\ref{fig:stableXtriplet}a). 

\subsection{Free space radiative quantum yield\label{paragraph:QY}}
\begin{figure}[!ht]
    %\centering
    \includegraphics{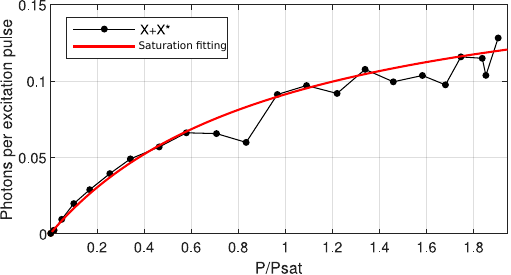}
    \caption{Saturation curve of the free space pQD2 emission, where the sum of the exciton and trion intensities is plotted as a function of the pulsed laser excitation power in unit of $P_{\text{sat}}$, with $P_{\text{sat}}= 220$~\si{\micro\watt}. The intensities are given in photons per pulse after correction for the overall collection efficiency of the setup $\eta_{col}\simeq0.1$. The solid line is a fit of the saturation curve, giving the pQD saturation flux $I_{\text{sat}}=0.18$~photon/pulse and hence the quantum yield of the emitter.}
    \label{fig:collectionSAT}
\end{figure}

The free space radiative quantum yield of a single photon emitter is assessed by the number of photons emitted per excitation pulse at saturation. The individual pQDs emit single photons at the excitonic transition as shown in Figure~2b of the main text, and their radiative quantum yield (or saturation flux) can thus be determined by measuring the PL intensity at saturation under pulsed excitation (using a frequency-doubled Ti:Sa laser with a repetition rate of 80~\si{\mega\Hz} and a temporal width of 130~\si{\femto\second}). We have previously shown that the pQDs exhibit alternating emission with the same radiative quantum yield for exciton and trion transitions due to the presence of photo-generated charges when the excitation power is increased (see sections~\ref{paragraph:XtrionXX} and ~\ref{paragraph:altXtrion}). To account for this effect and to not underestimate the radiative quantum yield of the pQDs, the sum of the exciton and trion emission intensities must be considered. In Figure~\ref{fig:collectionSAT}, the spectrally integrated emission shows a typical saturation behavior, which can be fitted by $I=I_{sat}\frac{P/P_{sat}}{1+P/P_{sat}}$, where $P$ is the excitation power, $P_{sat}$ the saturation power and $I_{sat}$ the saturation intensity. The detected intensity is converted into emitted photons per pulse by using the overall collection efficiency $\eta_{col}\simeq0.1$ (see section~\ref{paragr:colleff}). The radiative quantum yield is then directly given by the saturation plateau value $\eta_{\text{QY}}=0.18 \pm 0.02$. This radiative quantum yield is lower than that commonly measured for pQD ensembles in solution \cite{Protesescu} or thin films \cite{bo_perovskite_2021}, which generally exceed 50~\%. Such a difference can possibly be explained by the decrease in stability of the ligands passivating the pQD surface induced by the dilution and deposition processes, leading to additional non-radiative recombination channels.

\section{Collection efficiency \label{paragr:colleff}}
\begin{figure}[!ht]
    %\centering
    \includegraphics[width=0.99\textwidth]{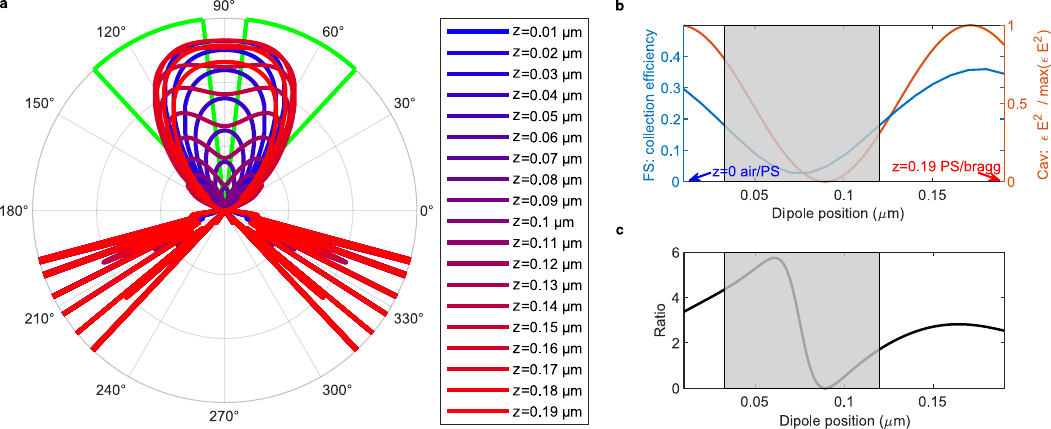}
    \caption{(a) Free space emission diagram calculated for a pQD characterized by three orthogonal dipoles at various positions $z$ within a 200~\si{\nano \metre}-thick layer of polystyrene (PS) deposited on a dielectric mirror. The position $z=0$ corresponds to the air/PS interface and $z=0.20$~\si{\micro \metre} to the PS/dielectric mirror interface. The green diagram indicates the angles collected by the drilled aspheric lens. The red pattern for angles larger than 180\si{\degree} corresponds to the emission into the substrate which is lost. (b)  Blue curve: Collection efficiency of the free space emission of a pQD as a function of its vertical position $z$. Orange curve: Evolution of the quantity $\epsilon(z) E(z)^2/\max(\epsilon E^2)$ (proportional to $g^2$, where $g$ is the vacuum Rabi coupling) as a function of the pQD position in the PS layer. The gray area corresponds to the positions where the collection efficiency is lower than half of the maximum. (c) Ratio of the orange and blue curves.}
    \label{fig:collection}
\end{figure}

In the free space configuration, numerical simulations \cite{reed_dipole_1987} were carried out to calculate the pQD emission diagram for different positions within the thickness of the polystyrene (PS) layer deposited on the planar dielectric mirror, as illustrated in Figure~\ref{fig:collection}a. Based on these simulations, the collection efficiency after the first lens was determined by considering the numerical aperture of the drilled aspherical lens ($\text{N.A.}=0.68$) and is displayed as the blue curve in Figure~\ref{fig:collection}b. For a 200~\si{\nano\metre}-thick PS layer, the first lens collection efficiency in free space ranges from 36~\si{\percent} when the pQD is in proximity to the PS/dielectric mirror interface to only 3~\si{\percent} when it is positioned closer to the air/PS interface. This variation is due to interference of the pQD emission with its reflection at the PS/dielectric mirror interface. 
Consequently, the vertical position of the pQD within the PS layer affects the estimation of the pQD radiative quantum yield $\eta_{\text{QY}}$ as well as the Purcell factor $F_p$. On the one hand, the variation of $F_p$ with the vertical position $z$ is given by $F_p=F_{p,\text{max}} \frac{\epsilon(z) E(z)^2}{\max(\epsilon E^2)}$, where $E$ is the electric field and $\epsilon$ the permittivity, and is depicted as the orange curve in Figure~\ref{fig:collection}b. On the other hand, as described in section~\ref{paragraph:QY}, the pQD radiative quantum yield is derived from the ratio of the saturation intensity to the collection efficiency such that $\eta_{\text{QY}}\propto \frac{I_{sat}}{\eta_{col}}$. The product $\eta_{\text{QY}} F_p$ is therefore proportional to the ratio of the orange and blue curves of Figure~\ref{fig:collection}b, as plotted in Figure~\ref{fig:collection}c. From an experimental point of view, the brightest pQDs are generally selected and are therefore more likely to be found in positions where the collection efficiency is high. We thus exclude the region where the collection efficiency is lower than half of the maximum, corresponding to the gray area in Figures~\ref{fig:collection}b and c. For the remaining positions, the ratio is relatively constant, showing that an incorrect estimate of the pQD vertical position will have only a very weak effect on the estimate of $\eta_{\text{QY}} F_p$. It is then reasonable to assume that a value of 36 \% can be used for the collection efficiency after the first lens. In any case, any potential bias in the estimate of this collection efficiency due to vertical positions variations would affect the coupling to the cavity in a similar manner, resulting in a negligible modification in the estimate of the Purcell factor. Taking into account the collection efficiencies of all the other optical elements in free space (spectrometer, camera, and transmission through optical elements), the overall detection efficiency in free space is 10~\si{\percent}.

\section{Time-resolved photoluminescence experiments}
\subsection{Wavelength dependence of the instrumental response function}
\begin{figure}[!ht]
    %\centering
    \includegraphics{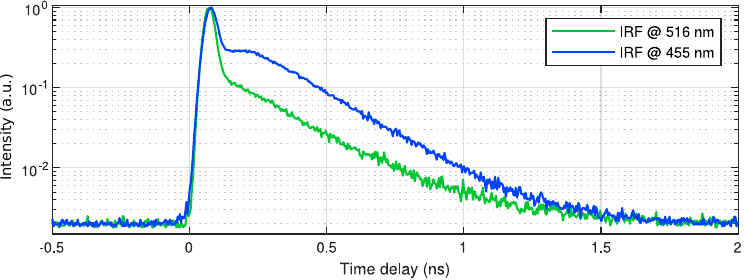}
    \caption{Instrumental Response Function (IRF) measured at the excitation wavelength (455~\si{\nano \metre}, blue line) and at a typical pQD emission wavelength (516~\si{\nano \metre}, green line).}
    \label{fig:IRFs}
\end{figure}

Given the short PL lifetime of pQDs ($\approx100$~\si{\pico\second}), which is further reduced by the Purcell effect in cavity, accounting for the Instrument Response Function (IRF) of the time-resolved PL setup is crucial when analyzing experimental data to accurately determine the PL lifetime. Figure~\ref{fig:IRFs} presents the IRF measured with a frequency-doubled 
tunable Ti:Sa pulsed laser of 80~\si{\mega\Hz} repetition rate and 130~\si{\femto\second} temporal width, at the excitation wavelength (455~\si{\nano \metre}) and at a typical pQD emission wavelength (516~\si{\nano \metre}). It shows a fast component of the order of 40~\si{\pico\second} and a long-lasting nanosecond tail which strongly depends on the wavelength, highlighting the importance of considering the IRF at the pQD emission wavelength for the PL decay analysis.  

\subsection{Photoluminescence decay analysis}
\begin{figure}[!ht]
    %\centering
    \includegraphics{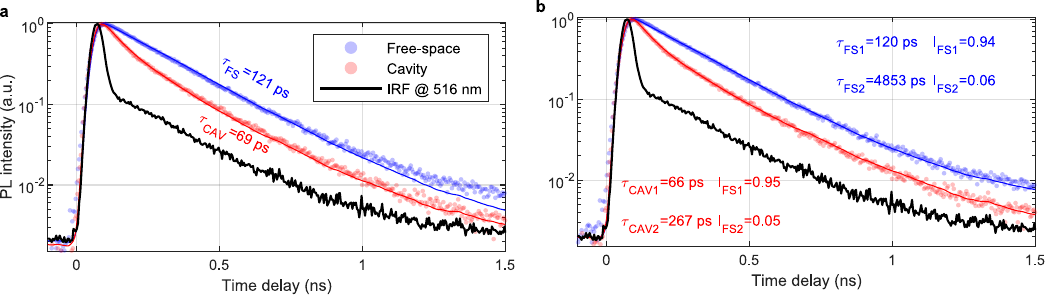}
    \caption{ (a) Time-resolved PL of pQD2 in free space (blue dots) and coupled to the cavity (red dots). The Instrument Response Function (IRF) is shown in black. The PL decays are fitted using a convolution of the IRF and a mono-exponential with lifetimes {$\tau_{\text{fs}}=121$~\si{\pico\second}} in free space (blue line) and $\tau_{\text{cav}}=69$~\si{\pico\second} in cavity (red line). (b) The same PL decays are fitted by a double exponential where 94\% of the intensity (the ratio being defined as $I=\frac{A_1 \tau_1}{A_1 \tau_1+A_2 \tau_2}$) decays with a lifetime $\tau_{\text{fs}}=120$~\si{\pico\second} in free space and 95\% of the intensity decays with a lifetime $\tau_{\text{cav}}=66$~\si{\pico\second} in cavity.}
    \label{fig:1expo2expos}
\end{figure}

To determine the acceleration of the pQD emission in the Purcell regime, we fit the experimental pQD PL decay in free space by:
\begin{equation}
            I_{\text{fs}}(t)=I_{\text{bckgd}}+\left(I_{\text{IRF}}*I_{\text{PL}}\right)(t),\qquad  \text{with}\qquad I_{\text{PL}}(t)=\Theta(t)Ae^{-t/\tau_{\text{fs}}},
        \label{eq:decays_fit_freespace}
\end{equation}
and in cavity by:
        \begin{equation}
            I_{\text{cav}}(t)=I_{\text{bckgd}}+\left(I_{\text{IRF}}*I_{\text{PL}}*I_{\text{stor}}\right)(t), \qquad \text{with}\qquad I_{\text{PL}}(t)=\Theta(t)Ae^{-t/\tau_{\text{cav}}} \qquad \text{and}\qquad I_{\text{stor}}=\Theta(t)e^{-t/\tau_{\text{stor}}},
        \label{eq:decays_fit_cavity}
        \end{equation}
where $*$ denotes a convolution product, $I_{\text{bckgd}}$ corresponds to a background signal from the detector dark counts, $I_{\text{IRF}}$ is the measured IRF at the pQD emission wavelength, $\Theta(t)$ is the Heaviside function, $A$ is an amplitude, and $\tau_{\text{fs}}$ and $\tau_{\text{cav}}$ are the pQD lifetime in free space and cavity configurations, respectively. For the PL decay of the cavity-coupled pQD, an additional convolution is needed to account for the cavity photon storage time $\tau_{\text{stor}}$, which is determined by the cavity finesse for each longitudinal order $p$ and ranges from 6 to 12~\si{\pico\second} (ignoring the photon storage contribution would lead to a larger acceleration factor). When analyzing the time-resolved PL data, $A$, $\tau_{\text{fs}}$ and $\tau_{\text{cav}}$ are free parameters. Figure~\ref{fig:1expo2expos}a, which is identical to Figure~2a in the main text, shows the PL decays in free space and in cavity, adjusted by a single exponential decay. A slight deviation of the fit can be observed at long time delays beyond 1~\si{\nano s}, which can be overcome by fitting the decay with a double exponential decay $I_{\text{PL}}(t)=\Theta(t)\left(A_1e^{-t/\tau_1}+A_2e^{-t/\tau_2}\right)$ (Figure~\ref{fig:1expo2expos}b). The associated additional long time can be attributed to the dark exciton contribution due to the thermal population transfer between the bright and dark exciton states \cite{amara_impact_2024}. Its contribution to the integrated intensity remains negligible and hardly changes the short time and the emission acceleration values.

\subsection{Polarization resolved photoluminescence decay of the excitonic doublet}
\begin{figure}[!ht]
    %\centering
    \includegraphics{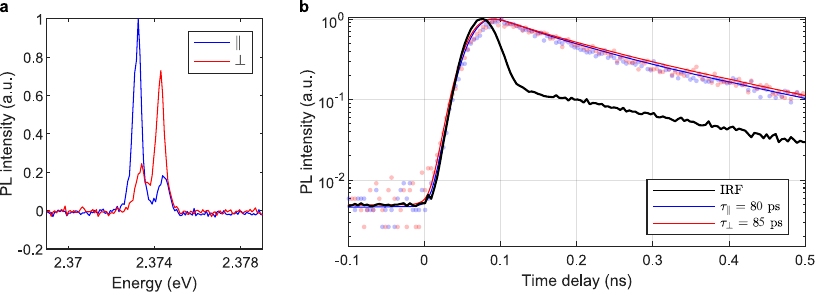}
    \caption{(a) Free space PL spectrum of the excitonic doublet of pQD4 detected in parallel (blue) and perpendicular (red) polarization. The parallel direction is defined for the orientation of the polarizer that maximizes the ratio of the blue line to the red line. (b) Time-resolved free space PL of pQD4 measured in parallel (blue dots) and perpendicular (red dots) polarization. The Instrument Response Function (IRF) is shown in black. The PL decays are fitted using a convolution of the IRF and a mono-exponential function (blue and red lines), yielding lifetimes {$\tau_{\text{$\parallel$}}=80 \pm 2$~\si{\pico\second}} and $\tau_{\text{$\perp$}}=85 \pm 2$~\si{\pico\second} in the parallel and perpendicular configurations, respectively.}
    \label{fig:TRPLPolar1Polar2}
\end{figure}

Free space time-resolved PL experiments were also performed on each excitonic emission line of a pQD exhibiting an emission doublet like the one studied in the main text. The two lines of the doublet can be separated in polarization in order to measure their lifetime separately (Figure~\ref{fig:TRPLPolar1Polar2}a). Since similar lifetimes are observed (Figure~\ref{fig:TRPLPolar1Polar2}b), we deduce that the two excitonic dipoles have the same emission lifetime, and that the PL decays can be performed by integrating the total emission of the pQD excitonic doublet without any loss of information.

\subsection{Statistics of the pQD emission acceleration in the Purcell regime}  
The cavity coupling was demonstrated on various individual pQDs along with their lifetime acceleration in the Purcell regime, as summarized in Table~\ref{table:pQDacceleration}.
\begin{table}[!ht]
%\centering
\renewcommand{\arraystretch}{1.5}\renewcommand{\arraystretch}{1.5}
\begin{tabular}{c||cccc||ccccc}
             & pQD1 & pQD2 & pQD3 & pQD4 & pQD5 & pQD6 & pQD7 & pQD8 & pQD9  \\ 
\hhline{=::====::=====}
$\tau_{\text{fs}}$ [\si{\pico\second}] & 63 & 121  & 66 & 61  & 92 & 156 & 47 & 82 & 134  \\
$\tau_{\text{cav}}$  [\si{\pico\second}]     & 39 & 69  & 38 & 40 &  44 & 109 & 37 & 50 & 84 \\
Acceleration             & 1.62   & 1.75 & 1.74 & 1.52 & 2.10 & 1.43 & 1.26 & 1.62 & 1.59 \\
Mode order $p$  & 14   & 17 & 16 & 14 & 16 & 16 & 14 & 16 & 14 \\
Figure \verb|#|  & 1,3,4,S5,S13,S14   & 2,S2,S3,S7,S10 & S4,S12 & S11 & / & / & / & / & /
\end{tabular}
\caption{\label{table:pQDacceleration}Example of the results obtained for different pQDs coupled to the fibered microcavity, showing the lifetime measured in free space $\tau_{\text{fs}}$ and in cavity $\tau_{\text{cav}}$, along with the corresponding acceleration factor defined as $\tau_{\text{fs}}/\tau_{\text{cav}}$, for a specific longitudinal mode order $p$. The numbers of the figures displaying experimental data for some of the studied pQDs are also listed.}
\end{table}

\subsection{Long term variation of the pQD lifetime in free space and in cavity}
\begin{figure}[!ht]
    %\centering
    \includegraphics{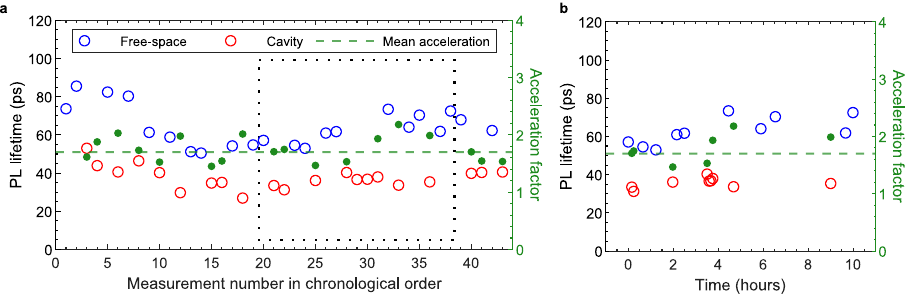}
    \caption{(a) PL lifetimes of pQD3 in free space (blue empty dots) and coupled to the cavity mode $p=16$ (red empty dots) as a function of the time order of the measurements. The emission acceleration defined as $\tau_{\text{fs}}/\tau_{\text{cav}}$ is shown for each cavity lifetime value (green dots), as well as the mean acceleration value $1.7\pm0.2$ (dashed green line). (b) PL lifetimes and accelerations as a function of time for measurements performed on the same day, corresponding to the dashed rectangle in (a).}
    \label{fig:chronologie_Tanger_pQD3}
\end{figure}

In general, all the individual pQDs we have studied show some variations over time of their free space emission lifetime of up to 50~\si{\percent}. Although the study of the origin of these variations is beyond the scope of this work, it is essential to take them into account when aiming to deduce a reliable value for the emission acceleration of a cavity-coupled pQD. To this end, we have successively measured the free space and cavity lifetimes of the same pQD several times. The results of these iterative measurements are summarized in Figure~\ref{fig:chronologie_Tanger_pQD3}a, together with the corresponding emission acceleration for each cavity measurement. We observe that the free space and cavity lifetimes vary consistently, resulting in a relatively constant emission acceleration with an average value of $1.7\pm0.2$ for this pQD. The typical time delay between free space and cavity measurements is on the order of tens of minutes. On this timescale, the free-space PL lifetime of the emitter remains stable, as shown in Figure~\ref{fig:chronologie_Tanger_pQD3}b, confirming that the acceleration is induced by the cavity.

\section{Theoretical framework}
\subsection{Emission spectrum of a two-level system coupled to a cavity}
The optical spectrum of the light exiting the cavity $S^{\text{X in cav}}(\omega)$ stems from a two-level system coupled to a cavity. It can be computed by solving the master equation followed by the use of the quantum regression theorem. We use an approach similar to the one in \cite{Auffeves2008} to which we added the pure dephasing contribution. We assume that, at the initial time, the two-level system is excited and the cavity is empty. This spectrum is given by:
\begin{equation}
S^{\text{X in cav}}(\omega,\omega_X,\omega_a) = \beta \ \frac{\frac{1}{|\omega-\tilde{\omega}_X|^2|\omega-\tilde{\omega}_a|^2}}{\int d\omega \frac{1}{|\omega-\tilde{\omega}_X|^2|\omega-\tilde{\omega}_a|^2}},
\label{eq:SXincav}
\end{equation}
where the complex effective energies of the emitter and the cavity mode are given by:
\begin{eqnarray}
\tilde{\omega}_X &\equiv& \omega_X + i \frac{\gamma+\gamma^*}{2} + g\ f\left(\frac{2 g}{\tilde{\delta}}\right) , \\
\tilde{\omega}_a &\equiv& \omega_a + i \frac{\kappa}{2} - g\ f\left(\frac{2 g}{\tilde{\delta}}\right) , \\
\tilde{\delta}&\equiv&\left(\omega_X+i\frac{\gamma+\gamma^*}{2}\right)-\left(\omega_a+i\frac{\kappa}{2}\right),
\end{eqnarray}
$\tilde{\omega}_X$ and $\tilde{\omega}_a$ are the roots of equation $\big(\omega-\omega_X-i \frac{\gamma+\gamma^*}{2}\big)\big(\omega-\omega_a-i \frac{\kappa}{2}\big)-g^2=0$ , 
where $\hbar \omega_X$ is the uncoupled two-level system energy transition, and $\hbar \omega_a$ the bare cavity resonant energy. $\gamma$ and $\kappa$ are respectively the emitter and the cavity decay rates (uncoupled), $\gamma^*$ is the pure dephasing rate and $g$ is the vacuum Rabi coupling strength. The function $f$ is defined as $f(z)\equiv \frac{\sqrt{1+z^2}-1}{z}$. The single photon efficiency $\beta$ is given by the probability of a photon to exit the cavity mode for each quantum of excitation in the two-level system and reads:
\begin{equation}
\beta = \frac{4 g^2}{4 g^2\left(1+\frac{\gamma}{\kappa}\right)+\gamma (\gamma+\gamma^*+\kappa)\left(1+\left(\frac{\omega_X-\omega_a}{(\gamma+\gamma^*+\kappa)/2}\right)^2\right)}.
\end{equation}
We underline that the expression of the spectrum $S^{\text{X in cav}}(\omega)$ is normalized such that its integral over $\omega$ gives the single photon efficiency.
This normalisation of the spectrum can be computed and reads:
\begin{equation}
\int_{-\infty}^{\infty}  d\omega \frac{1}{|\omega-\tilde{\omega}_X|^2 |\omega-\tilde{\omega}_a|^2} =\left(\frac{\gamma+\gamma^*+\kappa}{2\pi}\left[g^2+\frac{1}{4}(\gamma+\gamma^*)\kappa\left(1+\frac{4(\omega_X-\omega_a)^2}{(\gamma+\gamma^*+\kappa)^2}\right)  \right]  \right)^{-1} .
\end{equation}

\subsection{Spectral envelope in case of cavity modulation}
When the cavity is modulated, the spectral envelope $E(\omega)$ is given by the sum of all the spectra for each emitter-cavity detuning, weighted by the presence probability $P(\omega_a)$ for a particular detuning, and reads:
\begin{equation}
E(\omega) \equiv \int_{-\infty}^{\infty} P(\omega_a)  S^{\text{X in cav}}_{\omega_a}(\omega) d\omega_a.   
\end{equation}
If the modulation has a spectrum much broader than the typical linewidth of the emitter-cavity spectrum, we can make the following approximation:
\begin{equation}
    E(\omega) \simeq P(\omega) \int_{-\infty}^{\infty}  S^{\text{X in cav}}_{\omega_a}(\omega) d\omega_a.
\end{equation}
The integral can then be computed and expressed as a sum of two Lorentzian lines:
\begin{equation}
\int_{-\infty}^{\infty}  S^{\text{X in cav}}_{\omega_a}(\omega) d\omega_a = \frac{A^+}{1+\left(\frac{\omega-\omega_X}{\ell^+/2}\right)^2}   +     \frac{A^-}{1+\left(\frac{\omega-\omega_X}{\ell^-/2}\right)^2}  , \label{eq:envelope_double_lorentz}
\end{equation}
where
\begin{eqnarray}
\ell^{\pm} &=& \frac{\gamma_{all}+\Gamma_{all}}{2} \pm \tilde{\kappa} , \label{eq:widthell}\\    
A^{\pm} &=& \frac{\pm 32 g^4 \gamma_{all} \kappa \gamma^*}{\gamma^2 \Gamma_{all} \tilde{\kappa}(\gamma_{all}+\Gamma_{all}\pm2\tilde{\kappa})(\gamma_{all}^2-\gamma_{all}\Gamma_{all}+2\Gamma_{all}\kappa\pm2\gamma_{all}\tilde{\kappa})}, \\
\Gamma_{all} &=& \sqrt{\gamma_{all}^2 + \frac{4 g^2 \gamma_{all}(\gamma+\kappa)}{\gamma \kappa}} ,\\
\gamma_{all}&=&\gamma+\gamma^*+\kappa, \\
\tilde{\kappa} &=& \sqrt{-4g^2 + \left(\frac{\Gamma_{all}-\gamma_{all}}{2}+\kappa\right)^2}.
\end{eqnarray}
Except from the assumption of a slowly varying presence probability $P(\omega_a)$, no other approximations have been made in order to obtain these formulae. They are therefore valid in both the weak and strong coupling regimes. 

\subsection{Global fitting procedure}
In this section, we describe the procedure used to fit the experimental data for the spectral envelope and time decay to extract the vacuum Rabi coupling $g$ and the relative weight of spectral diffusion and pure dephasing in the spectral linewidth, as briefly described in the main text. At the end we give approximations for the different expressions. 
%Note that, even if the conditions for using the approximations are not always fulfilled by the experimental case, the aim is to give the dominant contribution.

\subsubsection{Definition of the terms used}
We start by defining the normalized Lorentzian and Gaussian functions as:
\begin{equation}
    \mathcal{L}_{\Gamma,\omega_0}(\omega)\equiv \frac{2}{\pi \Gamma} \frac{1}{1+\left(\frac{\omega-\omega_0}{\Gamma/2}{}\right)^2} , \qquad
    \mathcal{G}_{\sigma,\omega_0}(\omega)\equiv \frac{1}{\sqrt{2\pi}\sigma}\exp\left(-\frac{(\omega-\omega_0)^2}{2 \sigma^2} \right) .
\end{equation}

In this section, we will use the following notations:
\begin{itemize}
    \item $B$: offset (corresponds to an experimental flat background noise);
    \item $A_1$ (resp. $A_2$): amplitude of peak 1 (resp. peak 2);
    \item $\Delta$: free space splitting between peak 1 and peak 2;
    \item $\omega_a$: instantaneous cavity energy;
    \item $\overline{\omega_{a}}$:  mean cavity energy; 
    \item $\omega_X$: instantaneous mean energy of peaks 1 and  2;
    \item $\overline{\omega_{X}}$: time averaged value of $\omega_X$; 
    \item $\sigma_{\text{SD}}$: standard deviation of the spectral diffusion;
    \item $\gamma^*_1$ and $\gamma^*_2$:  pure dephasing terms related to peak 1 and peak 2;
    \item $\gamma$: free space emitter decay rate (the same on both lines); 
    \item $\Gamma_i$ : emitter instantaneous linewidth: $\Gamma_i=\gamma+\gamma^*_i$; 
    \item $\delta$: instantaneous detuning defined with respect to the mean position of the two peaks such that $\delta=\omega_X-\omega_a$;
    \item $\kappa$: instantaneous cavity linewidth (extracted from finesse measurements).
\end{itemize}

\subsubsection{Relationship between spectral diffusion and pure dephasing}
\begin{figure}[!ht]
    \centering
    \includegraphics{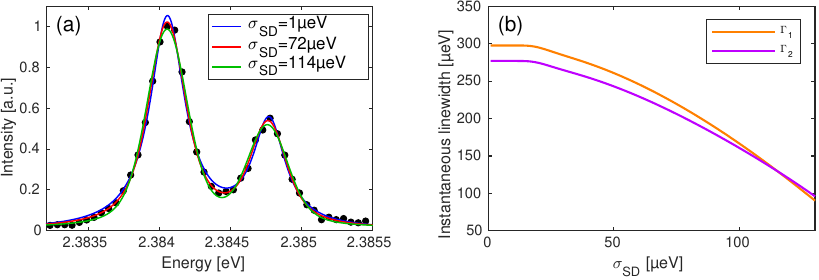}
    \caption{(a) Example of a free space PL spectrum together with the fits according to Eq. \ref{eq:fitFSSD} for three values of $\sigma_{\text{SD}}$. (b) Extracted instantaneous linewidths $\Gamma_1$  and $\Gamma_2$ as a function of $\sigma_{\text{SD}}$.}
    \label{fig:fsspectraSD}
\end{figure}

Here, the goal is to establish a relationship between the emitter instantaneous linewidth $\Gamma_i$ and the spectral diffusion standard deviation $\sigma_{\text{SD}}$.
The free-space emission spectrum is fitted to:
\begin{equation}
    A_1 (\mathcal{L}_{\Gamma_1,\omega_0-\Delta/2}*\mathcal{G}_{\sigma_{\text{SD}},0})(\omega) +A_2 (\mathcal{L}_{\Gamma_2,\omega_0+\Delta/2}*\mathcal{G}_{\sigma_{\text{SD}},0})(\omega) +B,
    \label{eq:fitFSSD}
\end{equation}
where $*$ stands for a convolution product. Despite the theoretical possibility of extracting independently the Gaussian component (which originates from spectral diffusion) and the Lorentzian one, the precision of this separation is compromised by noise. To circumvent this issue, the best instantaneous linewidths, i.e. $\Gamma_1$ and $\Gamma_2$, are extracted by fitting the spectra for each value of the spectral diffusion standard deviation $\sigma_{\text{SD}}$ (Figure~\ref{fig:fsspectraSD}(a)). The value of $\sigma_{\text{SD}}$ is then used as an input parameter. In Figure~\ref{fig:fsspectraSD}, the experimental PL spectrum is shown along with fits for three different values of $\sigma_{\text{SD}}$ ($1~\si{\micro eV}$, $72~\si{\micro eV}$ and $114~\si{\micro eV}$). The three curves are very similar, which explains why the contribution of spectral diffusion to the linewidth cannot be extracted directly from the fit. We note that  the sum of squared residuals, $\text{SSr}=\sum(y_{\text{data}}-y_{\text{fit}})^2$, reaches a minimum value for $\sigma_{\text{SD}}\simeq 70$~\si{\micro eV}, corresponding to $\hbar\Gamma_1\simeq225$~\si{\micro eV} and  $\hbar\Gamma_2\simeq210$~\si{\micro eV}. In the following, for each assigned value of $\sigma_{\text{SD}}$, the value of the instantaneous linewidth $\hbar\Gamma=\hbar(\gamma+\gamma^*)$ is given according to Figure~\ref{fig:fsspectraSD}(b).

\subsubsection{Vibration induced broadening of the cavity}
The mechanical vibrations of the cavity induce an effective broadening of the cavity peak, as shown in Figure~3a of the main text. This broadening is characterized by the standard deviation $\sigma_{\text{vib}}$, which depends on the longitudinal mode $p$ and is obtained by fitting the experimental data of the cavity transmission of a white lamp with:
\begin{equation}
    A\ (\mathcal{L}_{\kappa,\omega_0} * \mathcal{G}_{\sigma_{\text{vib}},0})(\omega) +B .
\end{equation}

\subsubsection{Fit of the spectral envelope}
In order to extract the Rabi coupling $g$, the experimental spectral envelope is fitted by:
\begin{equation}
    B+\iint d\omega_X d\omega_a\ \mathcal{G}_{\sigma_{\text{SD}},\overline{\omega_{X}}}(\omega_X)\ \mathcal{G}_{\sigma_{\text{vib}},\overline{\omega_{a}}}(\omega_a) \left(A_1 S^{\text{X in cav}}(\omega,\omega_X-\frac{\Delta}{2},\omega_a)+A_2 S^{\text{X in cav}}(\omega,\omega_X+\frac{\Delta}{2},\omega_a)\right),
\end{equation}
where $S^{\text{X in cav}}(\omega,\omega_X,\omega_a)$ is given by Eq.~\ref{eq:SXincav}. The pure dephasing term $\gamma^*$, needed to compute $S^{\text{X in cav}}$, is given by $\gamma^*_1=\Gamma_1-\gamma$ for the first peak of the doublet with a similar expression for the second. $\Gamma_1$ and $\Gamma_2$ are functions of the spectral diffusion width $\sigma_{\text{SD}}$ according to Figure~\ref{fig:fsspectraSD}(c). The value of $\sigma_{\text{vib}}$ is determined experimentally as described in the previous paragraph.  
 
For each value of the spectral diffusion $\sigma_{\text{SD}}$, the Rabi coupling strength $g$ is extracted. Finally, by combining decay and envelope analysis, the value of $g$ can be extracted along with the relative contribution of spectral diffusion and pure dephasing.

\subsubsection{Fit of the time decay}
Following the work by A. Auff\`eves \textit{et al.} \cite{Auffeves2010}, we define the effective emitter-cavity coupling rates for each peak of the excitonic doublet as: 
\begin{equation}
    R_{1,\delta}=\frac{4 g^2}{\kappa+\gamma+\gamma^*_1} \frac{1}{1+\left(\frac{\delta-\Delta/2}{(\kappa+\gamma+\gamma^*_1)/2}\right)^2}, \qquad R_{2,\delta}=\frac{4 g^2}{\kappa+\gamma+\gamma^*_2} \frac{1}{1+\left(\frac{\delta+\Delta/2}{(\kappa+\gamma+\gamma^*_2)/2}\right)^2},
\end{equation}
and the detuning dependant single photon efficiencies:
\begin{equation}
    \beta_{1,\delta}= \frac{R_{1,\delta} \kappa}{\gamma \kappa+R_{1,\delta}(\gamma+\kappa)}, \qquad \beta_{2,\delta}= \frac{R_{2,\delta} \kappa}{\gamma \kappa+R_{2,\delta}(\gamma+\kappa)}.
\end{equation}
For a particular detuning $\delta$, the time dependent PL decay is given by:
\begin{equation}
    I_{\text{short},\delta}(t)=\left(A_1 \beta_{1,\delta} (\gamma+R_{1,\delta}) e^{-(\gamma+R_{1,\delta})(t-t_0)} + A_2 \beta_{2,\delta} (\gamma+R_{2,\delta}) e^{-(\gamma+R_{2,\delta})(t-t_0)}\right) \Theta(t-t_0),
    \label{eq:Ishort}
\end{equation}
where $A_1$ and $A_2$ corresponds to the relative integrated amplitude of peak 1 and 2. $\Theta(t)$ is the Heaviside step function. For each peak of the doublet, the time integrated photoluminescence is proportional to the efficiency $\beta_i$. 

By summing the contribution of all detunings, we finally obtain the time dependent PL given by:
\begin{equation}
    I_{\text{short}}(t) = \int d\delta\ \mathcal{G}_{\sqrt{\sigma_{\text{SD}}^2+\sigma_{\text{vib}}^2},0}(\delta) \ I_{\text{short},\delta}(t).
\end{equation}
By considering a weak contribution of a long-time decay defined as:
\begin{equation}
    I_{\text{long}}(t)= A_{\text{long}} e^{-\gamma_{\text{long}}(t-t_0)} \Theta(t-t_0),
\end{equation}
as well as the storage of the emitted photons in the cavity before exiting the cavity, defined as:
\begin{equation}
    I_{\text{Stor}}(t) = e^{-\kappa (t-t_0)}\Theta(t-t_0),
\end{equation}
the experimental time-resolved decay is finally fitted by:
\begin{equation}
    (I_{\text{IRF}}*I_{\text{Stor}}*I_\text{short})(t) + (I_{\text{IRF}}*I_{\text{Stor}}*I_\text{long})(t), 
\end{equation}
where $*$ denotes a convolution product.
Similarly to the fit of the spectral envelope, the Rabi coupling strength $g$ is extracted for each value of $\sigma_{\text{SD}}$.

\subsubsection{Vacuum Rabi coupling}

\begin{figure}[!ht]
    \centering
    \includegraphics[width=1 \textwidth]{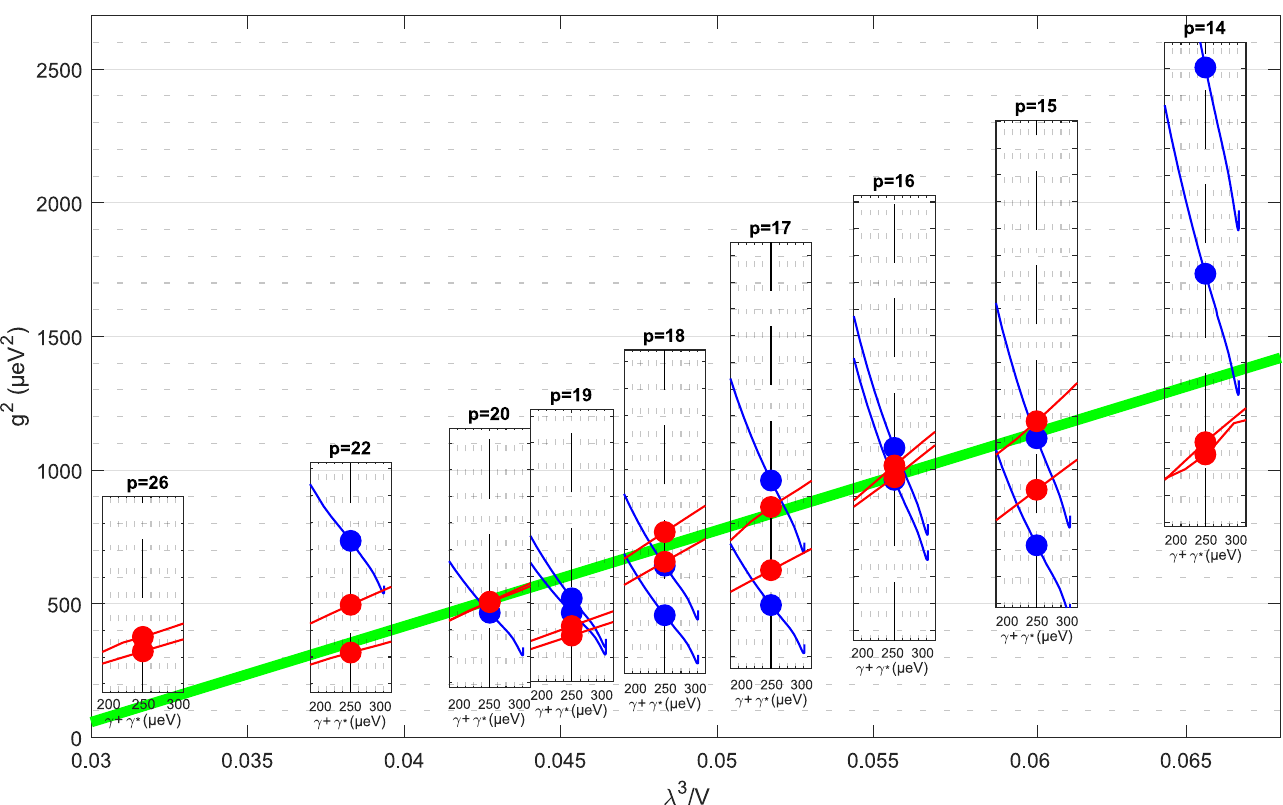}
    \caption{Vacuum Rabi coupling $g$ as a function of the instantaneous linewidth $\hbar(\gamma+\gamma^*)$ for different longitudinal cavity mode orders $p$. The blue curve is obtained from the analysis of the spectral envelope and the red one from the analysis of the PL decay. The longitudinal order $p=16$ corresponds to Figure~4a of the main text. }
    \label{fig:allcrossing}
\end{figure}

The procedure described in the previous paragraphs to extract the vacuum Rabi coupling $g$ was performed for several of longitudinal modes orders $p$. Figure~\ref{fig:allcrossing} shows the variation of $g$ extracted from the analysis of the PL decay and spectral envelope measurements as a function of the cavity mode order (except for $p=26$ where only PL decay measurements were performed). For almost all longitudinal orders, a crossing is observed for an instantaneous linewidth of $250 \pm50 $~\si{\micro\electronvolt}, which is then used to present the evolution of the Rabi coupling $g$ function of $\lambda^3/V$ for both spectral and temporal analysis shown in Figure~4b of the main text.  
%This gives a range of instantaneous linewidth where the Rabi coupling can  which is then used to plot the $g^2$ function of $\lambda^3/V$ Figure 4.b of the main text is $250~\si{\micro \electronvolt}$ .
%Note that for the lowest longitudinal mode order $p=14$, the crossing is not reached for an instantaneous linewidth smaller than the apparent emitter linewidth. This can be explained by the strongly reduced cavity vibrations for this contact mode, and thus a residual spectral mismatch between the emitter and the cavity could lead to a stronger effect on the determination of $g$. Indeed, we find that the $g$ value extracted from the spectral envelope analysis (blue curve) shows a much larger variation between two different measurements.

\subsection{Approximation of the theoretical expressions }
The aim of this section is to give a simpler understanding of why $g$ has opposite dependence with the instantaneous linewidth when extracted from decay or envelope analysis.
\subsubsection{Spectral envelope approximation}
In this paragraph, we give approximations of the spectral envelope expression Eq.~\ref{eq:envelope_double_lorentz} in order to emphasize the dominant dependence, although the requirements for these approximations are not always met in the experiments.

With our experimental parameters, we can estimate $\frac{g^2}{\kappa \gamma_{all}}\lesssim 0.05$ and $\frac{\gamma}{\gamma+\kappa}\simeq 0.05$. Assuming that $\frac{g^2}{\kappa \gamma_{all}} \ll 1$ and $\frac{\gamma}{\gamma+\kappa} \ll 1$, the widths of the two Lorentzian appearing in the envelope expression Eq.~\ref{eq:widthell} are approximately given by:
\begin{equation}
    \ell^- \simeq \gamma+\gamma^*, \qquad \qquad \ell^+ \simeq \kappa+\gamma_{all}\sqrt{1+\frac{4g^2(\gamma+\kappa)}{\gamma\kappa \gamma_{all}}}.
\end{equation}
The ratio of the two Lorentzian integrals reads:
\begin{equation}
    \frac{A^+\ell^+}{A^-\ell^-} \simeq \frac{\gamma+\gamma^*}{\kappa} \frac{\sqrt{1+\frac{4g^2(\gamma+\kappa)}{\gamma\kappa \gamma_{all}}}-1}{\sqrt{1+\frac{4g^2(\gamma+\kappa)}{\gamma\kappa \gamma_{all}}}+1}.
\end{equation}
Assuming that the coupling $g$ is small enough, i.e. $\frac{4g^2(\gamma+\kappa)}{\gamma\kappa \gamma_{all}} \ll 1$, the linewidths $\ell^\pm$ and the Lorentzian integral ratio are approximately given by: 
\begin{equation}
\ell^- \simeq \gamma+\gamma^*, \qquad \qquad \ell^+ \simeq \gamma+\gamma^*+2\kappa+\frac{2 g^2}{\gamma}, \qquad \qquad \frac{A^+\ell^+}{A^-\ell^-} \simeq \frac{\gamma^*}{\gamma^*+\kappa}\frac{g^2}{\kappa\gamma}. 
\end{equation}
We underline that the condition $\frac{4g^2(\gamma+\kappa)}{\gamma\kappa \gamma_{all}} \ll 1$ is not always fulfilled experimentally, since $\frac{4g^2(\gamma+\kappa)}{\gamma\kappa \gamma_{all}}$ can be as large as $\sim3.5$. The full model without approximations must therefore be used to analyze the experimental results.\\

Finally, we can approximate the spectral envelope with:
\begin{equation}
    E(\omega) \propto g^2 \left( S^{\text{fs}}(\omega) + \frac{\gamma^*}{\gamma^*+\kappa} \frac{g^2}{\kappa \gamma} \left(S^{\text{fs}}*\mathcal{L}_{2\kappa}\right)(\omega) \right),
\end{equation}
where $S^{\text{fs}}(\omega)$ is the free space emission spectrum, $\mathcal{L}_{2\kappa}$ stands for a normalized Lorentzian line shape with a $2\kappa$ width, and $*$ denotes a convolution product. For a vanishing coupling $g$, the envelope is simply the free space emission spectrum. When $g$ increases, the second term corresponds to the interaction between the emitter and the cavity. The spectral broadening by $2\kappa$, represented by $S^{\text{fs}}*\mathcal{L}_{2\kappa}$, can be understood as a resonance condition, where the limit of the resonance condition is reached when the cavity-emitter detuning equals $(\gamma+\gamma^*+\kappa)/2$. In that condition, the cavity emission occurs up to a detuning of $(\gamma+\gamma^*+\kappa)/2 + \kappa/2$ with respect to $\omega_X$, thus explaining the $2\kappa$ broadening.\\ 

\textbf{Inclusion of the spectral diffusion:}
\newline
In the previous approximate expression of the envelope, it is easy to introduce the effect of spectral diffusion. The relation between the instantaneous spectra (i.e. without the spectral diffusion) and the apparent spectra, including the broadening caused by spectral diffusion, is given by:
\begin{equation}
    S^{\text{fs}}_{\text{app}} = (S^{\text{fs}}_{\text{inst}}*P_{\text{SD}})(\omega),
\end{equation}
where the index "app" (resp. "inst") stands for apparent (resp. instantaneous) and $P_{\textbf{SD}}$ is the spectral diffusion energy distribution. Similarly, the relation between the instantaneous and the apparent envelope is given by:
\begin{equation}
    E_{\text{app.}} = (E_{\text{inst.}}*P_{\text{SD}})(\omega).
\end{equation}
Including the spectral diffusion thus leads to a similar approximated expression for the spectral envelope:
\begin{equation}
    E_{\text{app}}(\omega) \propto g^2 \left( S^{\text{fs}}_{\text{app}}(\omega) + \frac{\gamma^*}{\gamma^*+\kappa} \frac{g^2}{\kappa \gamma} \left(S^{\text{fs}}_{\text{app}}*\mathcal{L}_{2\kappa}\right)(\omega) \right).
    \label{eq:env_avec_SD}
\end{equation}
Here, we stress that the pure dephasing $\gamma^*$ can be different from the linewidth appearing in $S^{\text{fs}}_{\text{app}}(\omega)$ since it includes the broadening caused by spectral diffusion.\\

\textbf{Approximation of the dip:}
\newline
Considering the pQD-cavity platform as a superposition of two independent susbsystems, each consisting of a single two-level system coupled to a cavity mode, the envelope is the sum of two envelopes (with the approximate expression given by Eq.~\ref{eq:env_avec_SD}), one centered at $-\Delta/2$, and the other centered at $\Delta/2$. Assuming that $\gamma$, $\gamma^*$, $\kappa$ and $g$ are the same for both lines, we can add up the two envelopes, and thus the equation \ref{eq:env_avec_SD} is still valid. By assuming a symmetric spectrum, $S^{\text{fs}}_{\text{app}}(-\Delta/2)=S^{\text{fs}}_{\text{app}}(\Delta/2)$, the dip (as defined in the main text) is given by:
\begin{equation}
    dip \simeq \frac{S^{\text{fs}}_{\text{app}}(0) + \frac{\gamma^*}{\gamma^*+\kappa} \frac{g^2}{\kappa \gamma} \left(S^{\text{fs}}_{\text{app}}*\mathcal{L}_{2\kappa}\right)(0) }{S^{\text{fs}}_{\text{app}}(\frac{\Delta}{2}) + \frac{\gamma^*}{\gamma^*+\kappa} \frac{g^2}{\kappa \gamma} \left(S^{\text{fs}}_{\text{app}}*\mathcal{L}_{2\kappa}\right)(\frac{\Delta}{2}) }.
\end{equation}
 The normalized dip being finally given by:
\begin{equation}
    \frac{dip-dip_{fs}}{dip_{fs}} \simeq \frac{\gamma^*}{\gamma^*+\kappa} \frac{g^2}{\kappa \gamma}   \frac{\frac{ \left(S^{	\text{fs}}_{	\text{app}}*\mathcal{L}_{2\kappa}\right)(0)}{S^{	\text{fs}}_{	\text{app}}(0)}-\frac{ \left(S^{	\text{fs}}_{	\text{app}}*\mathcal{L}_{2\kappa}\right)(\frac{\Delta}{2})}{S^{	\text{fs}}_{	\text{app}}(\frac{\Delta}{2})}}{1+\frac{\gamma^*}{\gamma^*+\kappa} \frac{g^2}{\kappa \gamma}\frac{ \left(S^{	\text{fs}}_{	\text{app}}*\mathcal{L}_{2\kappa}\right)(\frac{\Delta}{2})}{S^{	\text{fs}}_{	\text{app}}(\frac{\Delta}{2})}}.
\end{equation}
We note that $\frac{ \left(S^{	\text{fs}}_{	\text{app}}*\mathcal{L}_{2\kappa}\right)(0)}{S^{	\text{fs}}_{	\text{app}}(0)}-\frac{ \left(S^{	\text{fs}}_{	\text{app}}*\mathcal{L}_{2\kappa}\right)(\frac{\Delta}{2})}{S^{	\text{fs}}_{	\text{app}}(\frac{\Delta}{2})}$ is positive with our experimental numerical values and depends only on apparent spectra, thus including the unknown contribution of spectral diffusion.

For sufficiently small $g$, where the denominator is approximately equal to $1$, we end up with:
\begin{equation}
    g^2 \propto  \left(1+\frac{\kappa}{\gamma^*}\right)\frac{dip-dip_{fs}}{dip_{fs}}. 
\end{equation}
For fixed experimental conditions, i.e. fixed dip values and fixed cavity linewidth $\kappa$, but unknown spectral diffusion, $g^2$ increases with $1/\gamma^*$.

\subsubsection{Time decay approximation}
In this section, the multiple exponential decays caused by various emitter-cavity detunings is approximated as a single effective exponential decay.\\

\textbf{Best single exponential decay approximation:}
\newline
We assume that a time signal is given by the sum of multiple exponential decays: 
\begin{equation}
    S(t) = \int C(x) e^{-\gamma(x) t}  dx.
\end{equation}
Finding the best effective single exponential decay $S_{\text{eff}}(t)=C_{\text{eff}}e^{-\gamma_{\text{eff}}t}$ that approaches the signal is equivalent to determining the best parameters $C_{\text{eff}}$ and $\gamma_{\text{eff}}$ that minimize the sum of the squared residuals given by:
\begin{equation}
    \chi= \int_0^\infty  \big(S(t)-S_{\text{eff}}(t)\big)^2 dt.
\end{equation}
The minimum of $\chi$ is reached when $\frac{\partial\chi}{\partial C_{\text{eff}}}=0$ and $\frac{\partial\chi}{\partial\gamma_{eff}}=0$, which implies:
\begin{equation}
    C_{\text{eff}} = \int dx\ C(x) \frac{2 \gamma_{\text{eff}}}{\gamma_{\text{eff}}+\gamma(x)} \qquad \text{and} \qquad C_{\text{eff}} = \int dx\ C(x) \frac{4 \gamma_{\text{eff}}^2}{(\gamma_{\text{eff}}+\gamma(x))^2}. \label{eq:Ceff}
\end{equation}
If we assume that $\gamma(x)$ does not vary too much, i.e. $\gamma(x) \simeq \gamma_{\text{eff}} (1+\epsilon(x))$, with $\epsilon(x)\ll 1$, then Eq.~\ref{eq:Ceff} leads to $C_\text{eff}=\int dx C(x)(1-\epsilon(x)/2)=\int dx C(x)(1-\epsilon(x))$, thus  $\int dx\ \epsilon(x) C(x) =0$, therefore:
\begin{equation}
    \gamma_{\text{eff}}= \frac{\int dx\ \gamma(x) C(x)}{\int dx\  C(x)}\qquad \text{and}  \qquad C_{\text{eff}}=\int dx\  C(x). 
    \label{eq:approxsingleexpo}
\end{equation}
\newline

\textbf{Single exponential decay approximation for an emitter in a varying detuning cavity:}
\newline
The formal expression for the decay in cavity is given by  Eq.~\ref{eq:Ishort}. Assuming $\gamma\ll\kappa$ leads to $\beta_i\simeq \frac{R_i}{R_i+\gamma}$. For simplicity, we can also assume that the two lines have the same amplitude, in which case the decay is given by:
\begin{equation}
    S(t) \simeq \int d\delta\ P_\delta \left(R_1 e^{-(\gamma+R_1)t}+ R_2 e^{-(\gamma+R_2)t} \right).
\end{equation}
The detuning distribution defined by $P_\delta$ includes both the effects of mechanical vibrations and spectral diffusion which play symmetrical roles. Using the relation given in Eq.~\ref{eq:approxsingleexpo}, the decay can be approximated by a single effective exponential decay:
\begin{equation}
    S(t) \simeq C e^{-(\gamma +R_{\text{eff}})t}, 
\end{equation}
where, if we assume that $P_\delta$ varies slowly (large modulation),
\begin{equation}
    R_{\text{eff}}\simeq \frac{\int d\delta \left(R_1^2+R_2^2 \right)}{\int d\delta \left(R_1+R_2 \right)} = \frac{2 g^2}{\kappa+\gamma^*}.
    \label{eq:Reffectif}
\end{equation}

Note that a very large modulation results in an effective acceleration that is exactly half that of the resonance case. We can therefore legitimately assume that the variation in decay between free space and cavity for any modulation will be given by $\zeta \frac{g^2}{\kappa+\gamma}$, with $\zeta$ being between 2 and 4.  

\subsubsection{Dominant dimensionless parameters}
From Eq.~\ref{eq:env_avec_SD} and Eq.~\ref{eq:Reffectif}, we can deduce that the dominant dimensionless parameters describing the spectral envelope and PL time decay (including the effect of mechanical vibrations and spectral diffusion) are given by:
\begin{eqnarray}
    \frac{\gamma^*}{\gamma^*+\kappa}\frac{g^2}{(\kappa+\gamma^*)\gamma}, & \qquad&\text{for the spectral envelope}, \\
    \frac{g^2}{(\kappa+\gamma^*)\gamma}, &\qquad &\text{for the time decay}.
\end{eqnarray}
                              
\maketitle

\nocite{*}
\bibliography{biblio_SI.bib}% Produces the bibliography via BibTeX.